\pdfoutput=1

\documentclass[prb,twocolumn,twoside,flushbottom,balancelastpage,superscriptaddress,showpacs,amsmath,amssymb,amsfonts]{revtex4}

\usepackage[final,notref,notcite]{showkeys} 
\usepackage{graphicx} 
\usepackage{color} 
\usepackage{amsmath} 
\usepackage{amssymb} 
\usepackage[squaren]{SIunits} 
\usepackage[sort&compress]{natbib} 
\usepackage{natmove} 
\usepackage[pdftitle={Elastic effects of vacancies in strontium titanate: Short- and long-range strain fields, elastic dipole tensors, and chemical strain},pdfauthor={Daniel A. Freedman, D. Roundy, T.A. Arias}]{hyperref} 
\usepackage{dcolumn} 
\usepackage{multirow} 
\usepackage{ifpdf} 
\usepackage[geometry]{ifsym} 
\usepackage{chemarrow} 
\usepackage{hyphenat} 

\ifpdf
\fi

\begin{document}

\title{Elastic effects of vacancies in strontium titanate:\texorpdfstring{\\}{ }Short- and long-range strain fields, elastic dipole tensors, and chemical strain}

\author{Daniel~A.~Freedman}
\affiliation{Laboratory of Atomic and Solid State Physics, Cornell University, Ithaca, New York 14853, USA}
\affiliation{Cornell Center for Materials Research, Cornell University, Ithaca, New York 14853, USA}
\author{D.~Roundy}
\affiliation{Department of Physics, Oregon State University, Corvallis, Oregon 97331, USA}
\author{T.A.~Arias}
\affiliation{Laboratory of Atomic and Solid State Physics, Cornell University, Ithaca, New York 14853, USA}
\affiliation{Cornell Center for Materials Research, Cornell University, Ithaca, New York 14853, USA}

\date{November 18, 2008}

\begin{abstract}
  We present a study of the local strain effects associated with
  vacancy defects in strontium titanate and report the first
  calculations of elastic dipole tensors and chemical strains for
  point defects in perovskites.  The combination of local and
  long\hyp{}range results will enable determination of
  x\nobreakdash-ray scattering signatures that can be compared with
  experiments.  We find that the oxygen vacancy possesses a special
  property --- a highly anisotropic elastic dipole tensor which almost
  vanishes upon averaging over all possible defect orientations.
  Moreover, through direct comparison with experimental measurements
  of chemical strain, we place constraints on the possible defects
  present in oxygen\hyp{}poor strontium titanate and introduce a
  conjecture regarding the nature of the predominant defect in
  strontium\hyp{}poor stoichiometries in samples grown via pulsed
  laser deposition.  Finally, during the review process, we learned of
  recent experimental data, from strontium titanate films deposited
  via molecular\hyp{}beam epitaxy, that show good agreement with our
  calculated value of the chemical strain associated with strontium
  vacancies.
\end{abstract}




\maketitle

\section{Introduction}
\label{sec:intro}

Perovskites in general, and strontium titanate in particular, are some
of the most frequently and exhaustively studied materials in
solid\hyp{}state physics and chemistry.  This attention has largely
derived from their diverse and interesting properties: high
piezoelectricity\cite{grupp1997gpe}, quantum
paraelectricity\cite{barrett1952dcp, muller1979siq},
ferroelectricity\cite{fleury1968spm, bednorz1984sax, gervais1987cos,
cohen1992ofp}, uniaxial stress\cite{uwe1976sif}, and colossal
magnetoresistance\cite{moritomo1996gmo, ramirez1997cm}.  Further, the
cubic perovskites manifest intriguing effects of underlying quantum
fluctuations, since, although they are comprised of relatively heavy
atomic constituents, a number of competing structures are
energetically and structurally similar\cite{zhong1994ptb}.
Characterization of the low\hyp{}temperature order parameters of these
materials remains an open and engaging question\cite{muller1991inp}.

Strontium titanate is a model perovskite --- commonly available and
reflective of many of the above properties of that material family.
Specifically, while strontium titanate is a wide band\hyp{}gap
insulator at room temperature, it exhibits semiconductivity at
elevated temperatures through doping or non\hyp{}stoichiometric
composition\cite{chan1981ns, eror1982htd} and superconductivity at low
temperatures through reduction via addition of
oxygen\cite{schooley1964sss, koonce1967stt}.  The structural phase
diagram of strontium titanate comprises a high\hyp{}temperature cubic
phase and a low\hyp{}temperature tetragonal phase, with a transition
temperature near $105$~K\cite{unoki1967esr, alefeld1969dmg,
vonwaldkirch1973fsn}.  The cubic perovskite structure is particularly
interesting due to the richness of its phase diagram (nonpolar
antiferrodistortive to ferroelectric to antiferroelectric
phases)\cite{zhong1995csi, lines1977paf}, to the capacity of these
phases to emerge from miniscule deviations from the cubic lattice and
its skeleton of octahedral oxygens (often through rigid rotations of
such), and to open questions regarding the types (displacive or
order\hyp{}disorder) of the transitions among these
phases\cite{zhong1994ptb}.

Defects and vacancies play a particularly important role in the
chemistry of perovskites and deserve continued study in strontium
titanate due to the electronic and superconducting effects of doping
as well as their role in the interface region of
heterostructures\cite{fleet2006mts}.  In the dilute limit, the
mechanics of defects are fully determined by an examination of
stress\hyp{}strain effects, in particular the elastic dipole tensor,
which motivates this work's emphasis on such a quantity.  Our
presentation of both short\hyp{}range displacements around a point
defect as well as long\hyp{}range effects (characterized by the
elastic dipole tensor) allows for the calculation of x\nobreakdash-ray
scattering signatures.  These quantities also enable the prediction of
defect mechanics, such as the behavior of defects within an externally
imposed strain gradient (as present in heterostructures), as well as
the ratio of chemical strain to stoichiometric deviation (a direct
experimental observable).  Finally, through comparison of our
predictions of chemical strain with experimental results, we draw a
number of important conclusions regarding the nature of point defects
in non\hyp{}stoichiometric strontium titanate.

\section{Background}
\label{sec:background}

Point defects introduce lattice distortions on both local and
long\hyp{}range scales.  While the short\hyp{}range distortions must
be described by a potentially large set of atomic displacements, the
long\hyp{}range \emph{elastic} distortions may be completely described
by a single tensor, the elastic dipole tensor\cite{gillan1984edt}.

The elastic dipole tensor and its relation to elastic effects may be
understood by the following simple considerations.  Consider to
quadratic order the most general expansion of the free energy per unit
volume of a crystal in terms of the strain $\epsilon_{ij}$ ($i$ and
$j$ refer to coordinate axes) and the number of defects per unit
volume $n_d$,
\begin{multline} \label{eqn:freeEnergy}
  f \left( \epsilon_{ij},n_d \right) = f_{0} + n_d \, E_d + \tfrac{1}{2} \, n_d^2 \, E_{dd}\\
  + \tfrac{1}{2} \sum_{ijkl} C_{ijkl} \, \epsilon_{ij} \, \epsilon_{kl} - n_d \sum_{ij} \epsilon_{ij} \, G_{ij}.
\end{multline}
The Taylor expansion coefficients $E_d$, $E_{dd}$, $C_{ijkl}$, and
$G_{ij}$ are, respectively, the defect formation energy, an average
inter\hyp{}defect interaction energy, the components of the
\emph{elastic stiffness tensor} of the material, and the components of
the \emph{elastic dipole tensor} of the defects.  The negative
derivative of the free energy \eqref{eqn:freeEnergy} with respect to
strain then gives the stress,
\begin{equation} \label{eqn:sigma}
  -\frac{\partial f}{\partial \epsilon_{ij}} \equiv \sigma_{ij} = - \sum_{kl} C_{ijkl} \, \epsilon_{kl} + n_d \, G_{ij}.
\end{equation}

To isolate the elastic dipole tensor, we can consider the rate of
change of the stress in the crystal per unit concentration of defects,
while holding \emph{strain fixed}, that is, under \emph{strain
control}.  Although challenging experimentally, strain control is
quite convenient computationally since it corresponds to performing
calculations with fixed lattice vectors.  This derivative thus gives
the elastic dipole tensor $\mathbf{G}$ directly,
\begin{equation} \label{eqn:g}
  \left. \frac{\partial \, \sigma_{ij}}{\partial \, n_d} \right|_{\boldsymbol{\epsilon}} \negthickspace = \, G_{ij},
\end{equation}
so that \emph{positive} diagonal components of $\mathbf{G}$ indicate
that the presence of defects tends to \emph{expand} the crystal along
the corresponding directions.

Alternatively, we can also consider the derivative of the strain in
the crystal per unit concentration of defects under \emph{stress} or
\emph{load control} (holding stress fixed).  While stress control is
computationally more complicated than strain control, it is the most
common experimental situation.  Under experimental conditions, the
crystalline lattice vectors relax such that there is essentially zero
stress (under normal laboratory conditions, atmospheric pressure
corresponds to a negligible stress).  This criterion allows the
relation of strain to a newly defined quantity,
$\boldsymbol{\Lambda}$,
\begin{equation} \label{eqn:lambda}
  \left. \frac{\partial \, \epsilon_{ij}}{\partial \, n_d} \right|_{\boldsymbol{\sigma}} \negthickspace = \sum_{kl} S_{ijkl} \, G_{kl} \, \equiv \, \Lambda_{ij},
\end{equation}
where $\mathbf{S}$, the \emph{elastic compliance tensor}, is the
inverse of the elastic stiffness tensor $\mathbf{C}$, and
$\boldsymbol{\Lambda}$ is defined as the \emph{defect\hyp{}strain
tensor}, which is the strain per unit defect concentration induced in
a crystal at fixed stress.

Relying upon the above derivations, the numerical calculation of the
elastic dipole tensor $\mathbf{G}$ is straightforward.  We compute the
stress induced with the introduction of a single defect in a supercell
(maintaining fixed lattice vectors, but allowing relaxation of the
atomic coordinates).  From Equation~\eqref{eqn:g}, this yields
\begin{align} \label{eqn:calcG}
  G_{ij} &= \tfrac{1}{n_d} \left( \sigma^{\,d}_{ij} - \sigma_{ij} \right)\\
         &= V_{\circ} \: \Delta \, \sigma_{ij}, \nonumber
\end{align}
where $\sigma^{\,d}$ and $\sigma$ are the stresses with and without
the defect in the cell, respectively, and $V_\circ$ is the supercell
volume.  Once $\mathbf{G}$ is known, $\boldsymbol{\Lambda}$ may also
be computed directly from Equation~\eqref{eqn:lambda}.  As a practical
matter, we note that in this approach, the lattice vectors need not be
those of a fully relaxed bulk crystal, provided the strain is small
and kept fixed.

Experimental works often report the variations in \emph{chemical
strain}, the strain due to the presence of defects, with respect to
stoichiometric deviations in the crystalline chemical formula.  From
the above considerations, the chemical strain is
$\boldsymbol{\epsilon}_c \equiv n_d \boldsymbol{\Lambda}$.  Deviations
in stoichiometry specify the number of defects per chemical unit,
$\delta$, so that, in this context, the concentration of defects per
unit volume is $n_d = \delta / V_c$, where $V_c$ is the volume of the
chemical unit.  These two relations then immediately provide the
chemical strain as proportional to this \emph{stoichiometric defect
deviation}, $\delta$,
\begin{equation} \label{eqn:chemStrain}
  \boldsymbol{\epsilon}_c = \left( \frac{\boldsymbol{\Lambda}}{V_c} \right) \delta.
\end{equation}
Experimentally, one does not generally obtain a full tensor for
$\boldsymbol{\epsilon}_c$ but, instead, an average over all equivalent
defect orientations which restore the symmetry of the underlying
crystal.  In cubic crystals, one measures a scalar $\epsilon_c$ which
corresponds to the mean diagonal component of
$\boldsymbol{\epsilon}_c$.

\section{Methods}
\label{sec:methods}

To simulate strontium titanate, we employ a shell\hyp{}potential
model\cite{dick1958tdc} parameterized for strontium
titanate\cite{akhtar1995css}.  Shell\hyp{}potential models are
formulated as an extension to ionic pair potentials and employed to
capture the polarizability of the atomic constituents.  The shell
model separates each ion into two parts, a core and an outer shell,
which possess individual charges that sum to the nominal charge of the
ion.  The total model potential $U$ consists of three terms,
\begin{equation} \label{eqn:u}
  U \equiv U_{\mspace{-1mu}P} + U_{C} + U_{\mspace{-1mu}B},
\end{equation}
representing, respectively, the polarizability of the ions and the
Coulomb and short\hyp{}range interactions among the ions.  The
polarizability is captured by harmonic springs connecting the core and
shell of each ion, so that $U_{\mspace{-1mu}P}$ has the form,
\begin{equation} \label{eqn:uP}
  U_{\mspace{-1mu}P} = \tfrac{1}{2} \sum_{i} \mspace{1mu} k_i \left|\Delta \mspace{1mu} r_i \right|^2,
\end{equation}
where $|\Delta \mspace{1mu} r_i|$ is the core\hyp{}shell separation
for ion $i$ and the $k_i$ are a set of ion\hyp{}specific spring
constants.  The Coulomb contributions take the form,
\begin{equation} \label{eqn:uC}
  U_{C} = \tfrac{1}{2} \sideset{}{'}\sum_{i,j} \frac{k_c q_i q_j}{r_{ij}},
\end{equation}
where $i$ and $j$ range over all cores and shells (excluding terms
where $i$ and $j$ refer to the same ion), $q_i$ and $q_j$ are the
corresponding charges, $r_{ij}$ is the distance between the charge
centers, and $k_c$ is Coulomb's constant.  We compute this Coulombic
interaction\cite{ewald1921boe} using a Particle Mesh Ewald
algorithm\cite{darden1993pme, essmann1995asp, deserno1998htm} with all
real\hyp{}space pair\hyp{}potential terms computed out to a fixed
cutoff distance using neighbor tables.  Finally, the short\hyp{}range
interactions are included through a sum of
Buckingham\cite{buckingham1958tri} pair potentials (which can be
viewed as combinations of Born\hyp{}Mayer\cite{born1932gi} and
Lennard\hyp{}Jones\cite{lennardjones1931c} potentials) of the form,
\begin{equation} \label{eqn:uB}
  U_{\mspace{-1mu}B} = \tfrac{1}{2} \sum_{i,j} \left( A_{ij} \mspace{1mu} e^{-r_{ij} / \rho_{ij}} - C_{ij} \mspace{1mu} r_{ij}^{-6} \right),
\end{equation}
where $i$ and $j$ range over \emph{all shells} and $A_{ij}$,
$\rho_{ij}$, and $C_{ij}$ are pair\hyp{}specific adjustable
parameters.  Here, the first term (Born\hyp{}Mayer) serves as a
repulsive short\hyp{}range interaction to respect the Pauli exclusion
principle, and the second term (Lennard\hyp{}Jones) models the
dispersion or van der Waals interactions\cite{vanderwaals1873odc}.
The specific electrostatic and short\hyp{}range shell\hyp{}model
parameters used in this study were fit to strontium titanate by Akhtar
et al.\cite{akhtar1995css}, with values as listed in
Tables~\ref{tab:potES} and~\ref{tab:potSR}.  Finally, we wish to
emphasize again, as it is rarely mentioned explicitly in the
shell\hyp{}potential literature, that the pair\hyp{}potential terms in
$U_{\mspace{-1mu}B}$ apply to the \emph{shells only}, and \emph{not}
to the cores.

\begin{table}
  \setlength{\doublerulesep}{0\doublerulesep}
  \setlength{\tabcolsep}{4.9\tabcolsep} 
  \begin{tabular}{cD{.}{.}{3.3}D{.}{.}{2.3}D{.}{.}{5.3}}
    \hline\hline\\[-1.5ex]
    \multirow{2}{*}{Ion} & \multicolumn{1}{c}{Shell} & \multicolumn{1}{c}{Core} & \multicolumn{1}{c}{Spring Constant}\\
    & \multicolumn{1}{c}{Charge [e]} & \multicolumn{1}{c}{Charge [e]} & \multicolumn{1}{c}{[eV$\cdot$\AA$^{-2}$]}\\[0.5ex]
    \hline\\[-1.5ex]
    Sr$^{2+}$ &   1.526 &  0.474 &    11.406\\
    Ti$^{4+}$ & -35.863 & 39.863 & 65974.0\\
    O$^{2-}$  &  -2.389 &  0.389 &    18.41\\[0.5ex]
    \hline\hline
  \end{tabular}
  \caption[Electrostatic shell\hyp{}model potential parameters for
    strontium titanate]{Electrostatic shell\hyp{}model potential
    parameters for strontium titanate (from Akhtar et
    al.\cite{akhtar1995css}).}
  \label{tab:potES}
\end{table}
\begin{table}
  \setlength{\doublerulesep}{0\doublerulesep}
  \setlength{\tabcolsep}{5\tabcolsep}
  \begin{tabular}{rD{.}{.}{5.2}D{.}{.}{1.5}D{.}{.}{2.1}}
    \hline\hline\\[-1.5ex]
    \multicolumn{1}{c}{Interaction} & \multicolumn{1}{c}{A [eV]} & \multicolumn{1}{c}{$\rho$ [\AA]} & \multicolumn{1}{c}{C [eV$\cdot$\AA$^6$]}\\[0.5ex]
    \hline\\[-1.5ex]
    Sr$^{2+}$ $\Leftrightarrow$ O$^{2-}$ &   776.84 & 0.35867 &  0.0\\
    Ti$^{4+}$ $\Leftrightarrow$ O$^{2-}$ &   877.20 & 0.38096 &  9.0\\
    O$^{2-}$ $\Leftrightarrow$ O$^{2-}$  & 22764.3  & 0.1490  & 43.0\\[0.5ex]
    \hline\hline
  \end{tabular}
  \caption[Short\hyp{}range shell\hyp{}model potential parameters for
    strontium titanate]{Short\hyp{}range shell\hyp{}model potential
    parameters for strontium titanate (from Akhtar et
    al.\cite{akhtar1995css}).}
  \label{tab:potSR}
\end{table}

Shell models have been extensively used for decades as the primary
empirical potential for modeling perovskites and other
oxides\cite{lewis1985pmf, catlow1983iis}.  We tested the correctness
of our coded implementation of this potential through comparisons of
lattice constants and elastic moduli and find excellent agreement.
For instance, using the same shell potential and ground\hyp{}state
structure, we predict a lattice constant for cubic strontium titanate
of $3.881$~\AA, which is within $0.3$\% of the value calculated by
Akhtar et al.\cite{akhtar1995css} For the elastic moduli, we find
$C_{11} = 306.9$~GPa, $C_{12} = 138.7$~GPa, and $C_{44} = 138.8$~GPa,
which are within $1.8$\%, $1.0$\%, and $0.7$\%, respectively, of the
values from Akhtar et al.\cite{akhtar1995css}.  From this, we conclude
that our implementation of the potential is correct.

We further note that the static dielectric constant of $216.99$, as
calculated by Akhtar et al.\cite{akhtar1995css} for the same parameter
set as our shell potential, is $28$\% lower than the experimental
value for strontium titanate of $301.00$, a level of agreement typical
of results from empirical potentials.  As Gillan\cite{gillan1981vfd,
gillan1983lrd} and Stoneham\cite{stoneham1983vcd} discuss, there
exists an important electrostriction effect whereby the tendency of
the crystal to screen electric fields impacts the elastic dipole
tensor.  (Note that this effect scales as $1-\varepsilon^{-1}$, where
$\varepsilon$ is the static dielectric constant\cite{gillan1983lrd}.)
Given the high dielectric constant of strontium titanate, this
screening is nearly perfect in both the actual experiment and our
model case; therefore, we expect that this effect is captured well in
our calculations below, despite our relatively large fractional error
in the static dielectric constant.

As is well known, strontium titanate has a large number of similar,
competing ground\hyp{}state structures\cite{zhong1995csi}.  We should
emphasize, at this point, that the main quantities of interest to this
study, either local atomic displacements or the elastic dipole and
defect\hyp{}strain tensors [from
Equations~\eqref{eqn:g}--\eqref{eqn:calcG}], are all defined as
defect\hyp{}induced changes relative to the bulk structure and so are
likely quite insensitive to which of the competing structures are used
to represent the bulk.

Accordingly, we have carried out what we regard as a thorough, but not
exhaustive, search for a \emph{probable} ground\hyp{}state structure.
Indeed, we have found no alternative structure which relaxes to an
energy less than our candidate ground\hyp{}state structure within our
potential.  We performed quenches on hundreds of random displacements
from the idealized positions of the $1 \times 1 \times 1$ primitive
unit cell to explore various potential reconstructions for supercells
up to $6\times 6 \times 6$.  We also considered a number of highly
ordered, human\hyp{}inspired configurations commensurate with the
antiferrodistortive disordering that is observed
experimentally\cite{abramov1995cba, muller1968csp} and predicted
theoretically\cite{pytte1969tsp, zhong1995csi, zhong1996eoq}.  Among
those minima which we explored, we selected the lowest\hyp{}energy
configuration to serve as the bulk crystalline state throughout this
study.  This configuration possesses a fairly regular pattern, namely,
each oxygen octahedron rotates slightly along $\pm[111]$ (trigonal)
directions in an alternating three\hyp{}dimensional $2 \times 2 \times
2$ checkerboard pattern.
\begin{figure}
  \begin{centering}
    \includegraphics[width=\columnwidth]{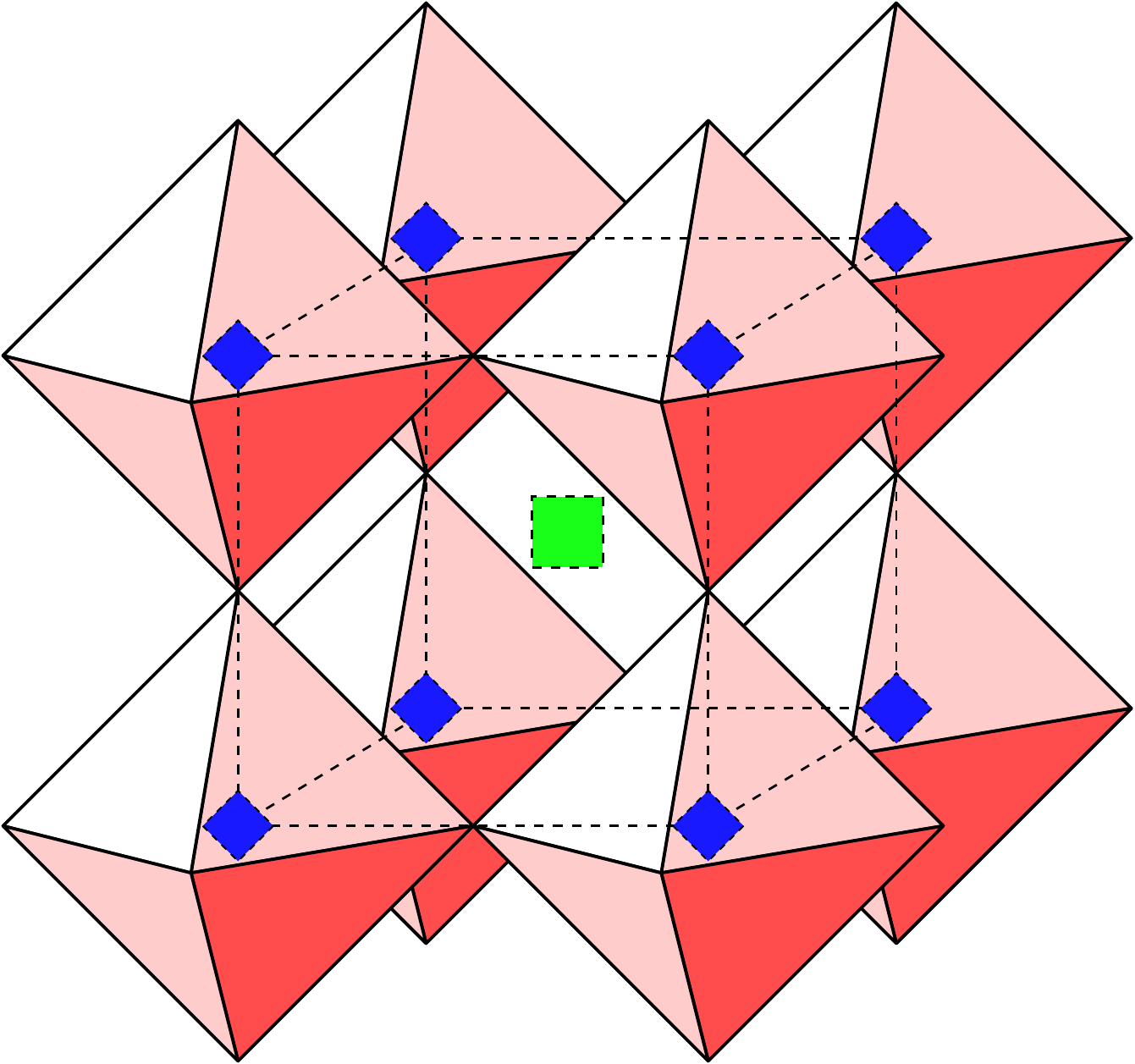}
    \caption[Atomic structure of strontium titanate]{Atomic structure
      of non\hyp{}reconstructed cubic strontium titanate: oxygen
      octahedra
      (\raisebox{-0.4ex}{\rlap{\rlap{\textcolor[rgb]{1.0,0.5,0.5}{\BigLowerDiamond}}{\textcolor[rgb]{1.0,0.5,0.5}{\BigRightDiamond}}}{\rlap{\rlap{\BigVBar}{\BigHBar}}{\BigDiamondshape}}})
      surrounding titanium atoms
      (\raisebox{-0.4ex}{\rlap{\textcolor[rgb]{0.1,0.1,1.0}{\FilledDiamondshape}}{\Diamondshape}}),
      with strontium atoms
      (\raisebox{-0.4ex}{\rlap{\textcolor[rgb]{0.1,1.0,0.1}{\FilledSquare}}{\Square}})
      outside the octahedra and equidistant from the titanium sites.}
    \label{fig:structureSTO}
  \end{centering}
\end{figure}
To aid in visualizing this reconstruction,
Figure~\ref{fig:structureSTO} depicts the atomic structure of
reference non\hyp{}reconstructed cubic strontium titanate, with the
oxygen octahedra lattice indicative of perovskite materials.

We investigated five defects in strontium titanate: the oxygen,
strontium, and titanium vacancies and the strontium\hyp{}oxygen and
titanium\hyp{}oxygen divacancies.  Since the octahedra rotations break
the original crystalline symmetry and generate a set of different
symmetry\hyp{}related reconstructions, each of these five defects can
be situated in multiple equivalent sites within each reconstruction.

Even well below room temperature, strontium titanate shows local
fluctuations among the possible reconstructions.  Thus, in addition to
raw results for a specific realization of each defect in a given
reconstruction, we also report results for each defect averaged over
all possible reconstructions for a given orientation of the defect.
The strontium and titanium vacancies do not select a specific
direction and, thus, this averaging represents the full cubic
crystalline symmetry group; their respective tensors therefore are
always diagonal with cubic symmetry.  The other defects do select
specific crystalline directions, which must be specified when
reporting the corresponding defect tensors.  We thus
reconstructionally average these latter defect tensors using either
symmetry arguments or explicit calculations with different
reconstructions, as appropriate.  Finally, with a view to chemical
strain measurements in macroscopic samples, we also report final
averages over all defect orientations.

In the case of the oxygen vacancy, the oxygen sits between two unique
nearest\hyp{}neighbor titanium atoms, thus uniquely distinguishing the
Ti\nobreakdash-V$_\text{\!O}$\nobreakdash-Ti direction, which we
define as $[100\mspace{1mu}]$, among the three cubic axes.  Next, the
titanium\hyp{}oxygen divacancy selects a unique
V$_\text{\!Ti}$\nobreakdash-V$_\text{\!O}$ direction, which we define
as $[\mspace{1mu}\bar100\mspace{1mu}]$, directed from the titanium
site toward the oxygen site.  Finally, the strontium\hyp{}oxygen
divacancy selects a unique V$_\text{\!Sr}$\nobreakdash-V$_\text{\!O}$
direction, which we define as $[1\bar10\mspace{1mu}]$, directed from
the strontium toward the oxygen site.  Once the reconstructional
averaging is accomplished, the average over defect orientations
requires generation of the crystal's response to all different
possible orientations (three for the oxygen vacancies, six for the
titanium\hyp{}oxygen divacancy, and twelve for the
strontium\hyp{}oxygen divacancy) and restores full cubic symmetry.

To obtain the results in Section~\ref{sec:results}, each of these
defects was placed within the bulk\hyp{}reconstructed strontium
titanate supercell, with cubic symmetry as experimentally observed for
$T \gtrsim 105$~K\cite{unoki1967esr, alefeld1969dmg,
vonwaldkirch1973fsn}, and then relaxed via the technique of
preconditioned conjugate gradient
min\-i\-mi\-za\-tion\cite{hestenes1952mcg} (specifically, the
Polak-Ribi\`{e}re\cite{polak1969ncm} method) to find the minimal
energy configuration (to within $\sim$$1$~\micro{}eV).  Supercell
convergence studies examined all five defects in cells containing up
to $13\,720$~atoms and verified that such defects were sufficiently
separated so that interactions across periodic supercell boundaries
were negligible\cite{leslie1985tee}.  The final relaxed atomic
configurations in the largest cells ($14 \times 14 \times 14$) provide
the local strain fields which we report below for each defect.  To
determine the long\hyp{}range strain fields, we compute the elastic
dipole tensor $\mathbf{G}$ through the prescription described in
Equation~\eqref{eqn:calcG} above; namely, we calculate the stress
induced by the introduction of a single defect in the supercell,
holding the lattice vectors fixed while allowing the atomic
coordinates to relax.

\section{Results}
\label{sec:results}

For a series of important and fundamental strontium titanate defects,
we examine both the elastic dipole and defect\hyp{}strain tensors, as
introduced in Section~\ref{sec:background}, as well as the local
strains surrounding each defect.  This section first examines the role
of the oxygen vacancy as a case study of a defect in strontium
titanate.  Subsequently, the results for the same set of studies are
presented for four other point defects in strontium titanate:
strontium and titanium vacancies and strontium\hyp{}oxygen and
titanium\hyp{}oxygen divacancies.  Section~\ref{sec:discussion}
continues with more general implications of our results.

\subsection{Oxygen vacancy}
As described above, there are three distinct orientations for the
oxygen vacancy as defined by the
Ti\nobreakdash-V$_\text{\!O}$\nobreakdash-Ti direction.  Moreover,
because of the reconstruction, there are in fact two distinct classes
of site within each possible orientation.  As one moves in the
positive sense of direction along the
Ti\nobreakdash-V$_\text{\!O}$\nobreakdash-Ti axis, these sites are
distinguished by whether the rotation of the octahedra surrounding the
two titanium sites changes from positive to negative or from negative
to positive (in the right\hyp{}hand sense about the $+[111]$
direction).  Below, we report results for the latter type of site.

We first examine the elastic dipole tensor computed according to
Equation~\eqref{eqn:calcG}.  To explore convergence, we compute the
dipole tensor components in supercells of sizes $2 \times 2 \times 2$,
$4 \times 4 \times 4$, $\ldots$ , $14 \times 14 \times 14$, containing
between $40$ and $13\,720$~atoms with defect separations between
$\sim$$8$ and $\sim$$54$~\AA.

\begin{figure}
  \begin{centering}
    \includegraphics[width=\columnwidth]{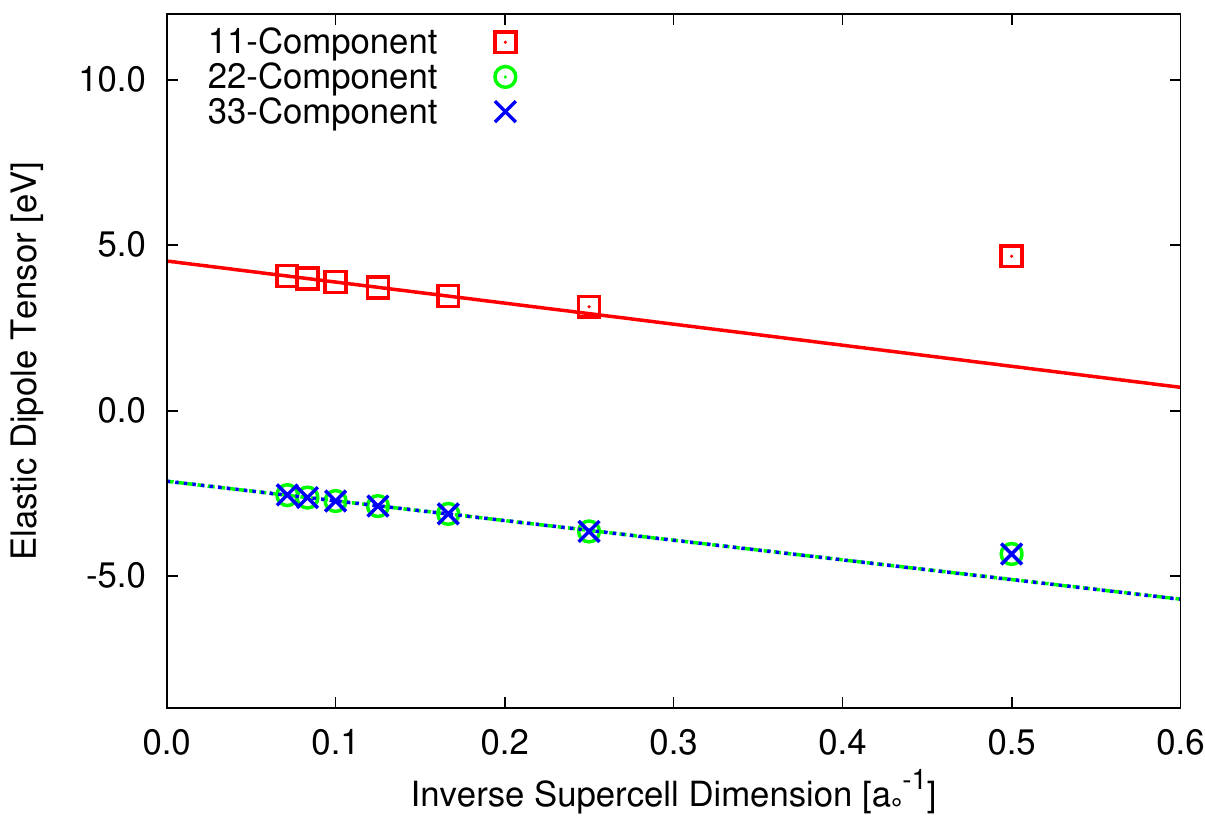}
    \caption[Elastic dipole tensor for oxygen vacancy in strontium
      titanate]{Diagonal components of elastic dipole tensor for
      oxygen vacancy in strontium titanate.  Data show linear
      convergence with inverse linear dimension of supercell size.}
    \label{fig:oElasticDipole}
  \end{centering}
\end{figure}

Figure~\ref{fig:oElasticDipole} depicts the convergence of the
diagonal components of the elastic dipole tensor as a function of
inverse linear dimension of the supercell.  The linear behavior in the
figure for large cells indicates that this quantity converges in the
same way as the Coulomb interaction between defects and allows
extrapolation of the fully converged value for these components in an
infinite supercell.  We observe exactly the same linear behavior with
inverse linear dimension of cell for the convergence of the
off\hyp{}diagonal components of the dipole tensor.  A linear fit to
the data, for all tensor components for cell sizes in the range
exhibiting linear behavior ($6 \times 6 \times 6$ through $14 \times
14 \times 14$), yields the elastic dipole tensor, \emph{extrapolated
to infinite cell size}, for this oxygen vacancy in strontium titanate,
\begin{equation*}
  \mathbf{G}_{\text{O}} = \left(
  \setlength{\arraycolsep}{2\arraycolsep}
  \begin{array}{D{.}{.}{2.2}D{.}{.}{2.2}D{.}{.}{2.2}}
     4.53 & -3.11 & -3.11\\
    -3.11 & -2.13 &  1.06\\
    -3.11 &  1.06 & -2.13\\
  \end{array}
  \right) \text{eV}.
\end{equation*}

As described above in Section~\ref{sec:methods}, at finite
temperatures ($T \gtrsim 105$~K), local fluctuations in the
reconstruction make it appropriate to average this tensor over all
reconstructions.  The result is to eliminate the off\hyp{}diagonal
components, leaving the reconstructionally averaged elastic dipole
tensor,
\begin{equation*}
  \overline{\mathbf{G}}_{\text{O}} = \left(
  \setlength{\arraycolsep}{2\arraycolsep}
  \begin{array}{D{.}{.}{1.2}D{.}{.}{2.2}D{.}{.}{2.2}}
    4.53 &  0.00 &  0.00\\
    0.00 & -2.13 &  0.00\\
    0.00 &  0.00 & -2.13\\
  \end{array}
  \right) \text{eV}.
\end{equation*}
Since the underlying, non\hyp{}defected, crystal is now cubic, we can
readily apply Equation~\eqref{eqn:lambda}
to $\overline{\mathbf{G}}_{\text{O}}$ to obtain the
reconstructionally averaged defect\hyp{}strain tensor,
\begin{equation*}
  \overline{\boldsymbol{\Lambda}}_{\text{O}} = \left(
  \setlength{\arraycolsep}{2\arraycolsep}
  \begin{array}{D{.}{.}{2.2}D{.}{.}{2.2}D{.}{.}{2.2}}
    16.33 &  0.00 &  0.00\\
     0.00 & -8.05 &  0.00\\
     0.00 &  0.00 & -8.05\\
  \end{array}
  \right) \text{\AA}^3.
\end{equation*}

The above results indicate that the oxygen vacancy tends to cause the
crystal to expand along the
Ti\nobreakdash-V$_\text{\!O}$\nobreakdash-Ti direction and to contract
along the two orthogonal directions by an amount which results in
negligible net volume change in the crystal.  When the above tensor is
averaged over all defect orientations (permutations of the three
coordinate axes), the result is near perfect cancellation, resulting
in a constant tensor (multiple of the identity) with a uniform
chemical strain per unit defect concentration of $+0.07$~\AA$^3$.
This corresponds to a ratio of chemical strain $\epsilon_c$ to the
deviation from stoichiometry $\delta$ in SrTiO$_{3-\delta}$ of
$\epsilon_c / \delta = +0.001$, indicating a \emph{very slight}
tendency for the crystal to expand due to the presence of oxygen
vacancies.

\begin{figure}
  \begin{centering}
    \includegraphics[width=\columnwidth]{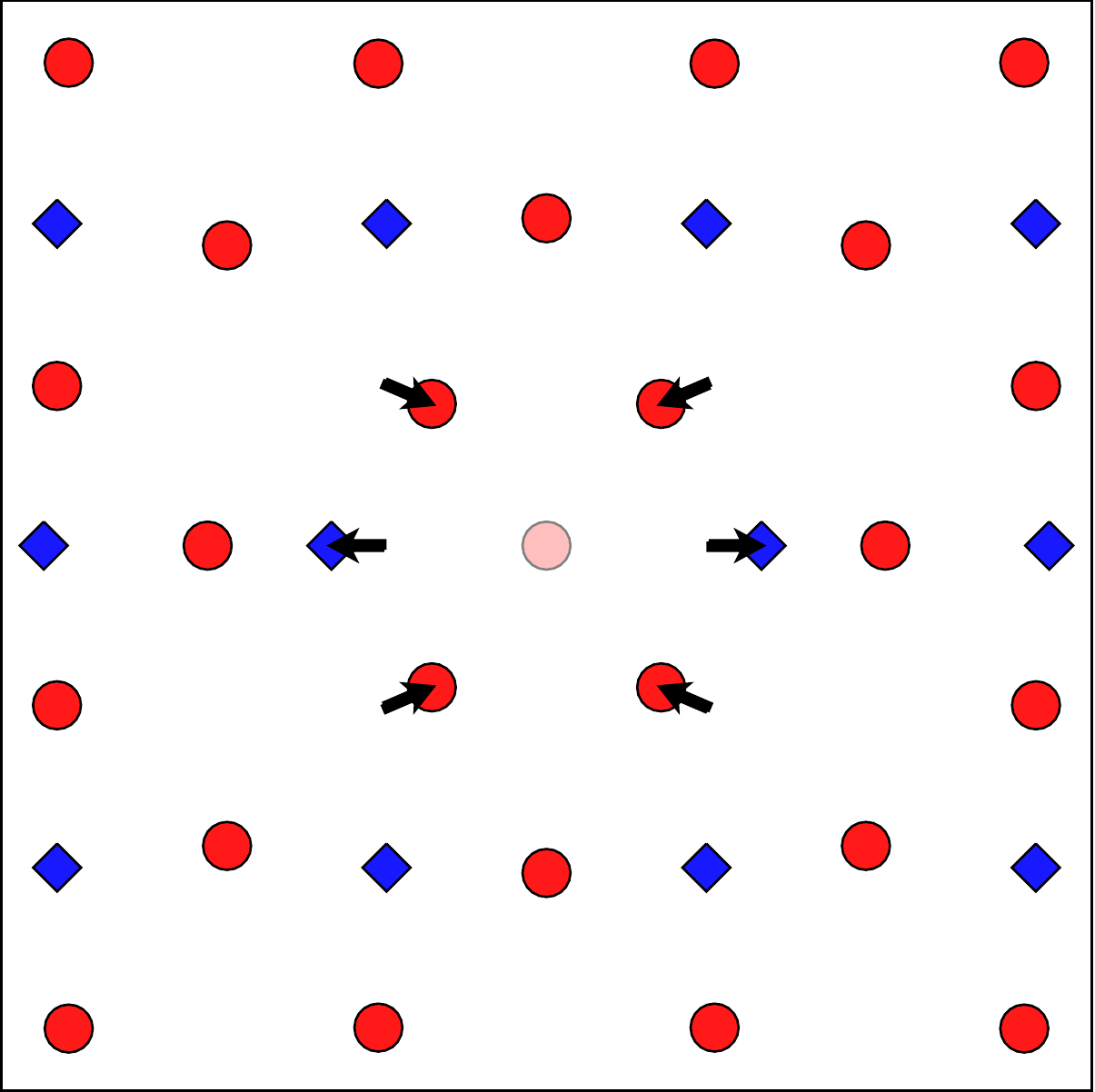}
    \caption[Local strain pattern for oxygen vacancy in strontium
      titanate]{Local strain pattern for oxygen vacancy in the
      $\hat{e}_1 \hat{e}_2$ plane of strontium titanate: titanium
      atoms
      (\raisebox{-0.4ex}{\rlap{\textcolor[rgb]{0.1,0.1,1.0}{\FilledDiamondshape}}{\Diamondshape}}),
      oxygen atoms
      (\raisebox{-0.4ex}{\rlap{\textcolor[rgb]{1.0,0.1,0.1}{\FilledCircle}}{\Circle}}),
      oxygen vacancy
      (\raisebox{-0.4ex}{\rlap{\textcolor[rgb]{1.0,0.75,0.75}{\FilledCircle}}{\textcolor[gray]{0.60}{\Circle}}}).
      Atomic displacements exaggerated by a factor of three for
      clarity (\pmb{\chemarrow}), displayed for significant
      in\hyp{}plane displacements ($> 0.1$~\AA) only.}
    \label{fig:oLocalStrain}
  \end{centering}
\end{figure}

Following Figure~\ref{fig:oLocalStrain}, we now examine the local
strain of this system after reconstructional averaging.  We note that
the removal of the oxygen ion, with a nominal charge of $-2$, leaves
an effective local positive environment in the location of the
vacancy.  We should expect pronounced Coulombic response to this
environment in the form of local crystal polarization.  Indeed, the
nearest neighbors of the oxygen vacancy, the two titanium atoms, now
move directly \emph{away} from the vacancy on the precise vector
connecting them, by $0.21$~\AA.  The next\hyp{}nearest neighbors are
the eight oxygens comprised of two equidistant rings of four oxygens
each, each of which move $0.21$~\AA\ \emph{toward} the vacancy
($0.19$~\AA\ along $\hat{e}_1$ with remaining projection of
$0.08$~\AA\ either along $\hat{e}_2$ or $\hat{e}_3$ as dictated by
symmetry), and one ring of four strontium atoms, each of which moves
$0.10$~\AA\ directly \emph{away} from the vacancy.  The third shell of
neighbors, a set of six oxygens, each one lattice constant away from
the vacancy along the three lattice directions in the crystal, moves
different amounts depending upon the vector; the two oxygens along the
Ti\nobreakdash-V$_{\text{\!O}}$\nobreakdash-Ti direction move
$0.06$~\AA\ directly \emph{away} from the vacancy (pushed sterically
by the second\hyp{}shell titanium atoms), while the other four oxygens
move by a negligible amount (only $0.01$~\AA).  While the fourth shell
of neighbors, a set of eight titanium atoms at a distance $\sqrt{5}/2
\, a_\circ$ from the vacancy, moves negligibly, the fifth shell of
neighbors (sixteen oxygen and eight equidistant strontium atoms) shows
significant movement in the strontium atoms of $0.15$~\AA\ \emph{away}
from the vacancy ($0.08$~\AA\ along $\hat{e}_1$ with remaining
projections of $0.09$~\AA\ along both $\hat{e}_2$ and $\hat{e}_3$),
even though the oxygen atoms move a negligible amount.  Finally, in
the sixth shell of neighbors, a set of twelve oxygen atoms, eight
oxygens (those not in the plane containing the defect and
perpendicular to $\hat{e}_1$) move $0.10$~\AA\ \emph{toward} the
vacancy ($0.02$~\AA\ along $\hat{e}_1$ with remaining projection of
$0.10$~\AA\ either along $\hat{e}_2$ or $\hat{e}_3$ as dictated by the
symmetry), while four others (those in the plane perpendicular to
$\hat{e}_1$) move a negligible amount.  All other atoms in the crystal
move less than $0.06$~\AA.

Finally, we would like to comment on the relation between local
strains and defect tensors.  We observe that the direction of motion
of the near\hyp{}neighboring atoms often correlates with the
far\hyp{}field motion described by the defect\hyp{}strain tensor.  In
this case, in the first and second shells, we see a general pattern of
movement which is away from the defect along $\hat{e}_1$ and toward
the defect in the other two directions, consistent with the signs in
the long\hyp{}range defect\hyp{}strain tensor.

\subsection{Strontium vacancy}
\begin{figure}
  \begin{centering}
    \includegraphics[width=\columnwidth]{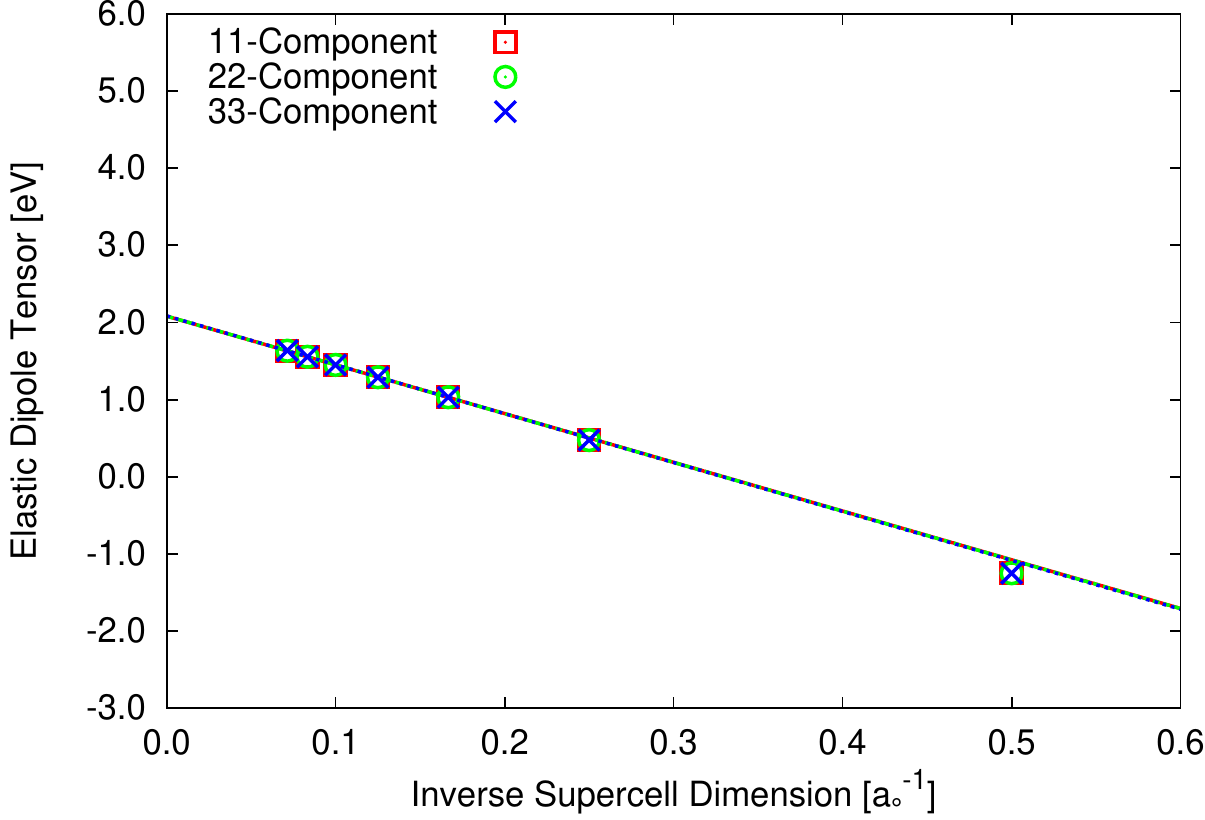}
    \caption[Elastic dipole tensor for strontium vacancy]{Diagonal
      components of elastic dipole tensor for strontium vacancy in
      strontium titanate.  Data show linear convergence with inverse
      linear dimension of supercell size.}
    \label{fig:srElasticDipole}
  \end{centering}
\end{figure}
We now repeat the above procedures to obtain similar results for the
strontium vacancy.  As described above, the strontium\hyp{}vacancy
site defines no unique direction and reconstructional averaging
recovers the full cubic symmetry group.  For any realization of the
reconstruction there are in fact two distinct types of strontium
sites.  Each such site sits at the center of a cube with oxygen
octahedra at the corners with alternating signs of rotations.  The
results reported below, prior to reconstructional averaging,
correspond to the site in which the rotation at the $[111]$ corner is
positive.

For the elastic dipole tensor we find
\begin{equation*}
  \mathbf{G}_{\text{Sr}} = \left(
  \setlength{\arraycolsep}{2\arraycolsep}
  \begin{array}{D{.}{.}{2.2}D{.}{.}{2.2}D{.}{.}{2.2}}
     2.08 & -0.23 & -0.23\\
    -0.23 &  2.08 & -0.23\\
    -0.23 & -0.23 &  2.08\\
  \end{array}
  \right) \text{eV},
\end{equation*}
where Figure~\ref{fig:srElasticDipole} shows the convergence of the
diagonal elements of the above tensor.
\begin{figure}
  \begin{centering}
    \includegraphics[width=\columnwidth]{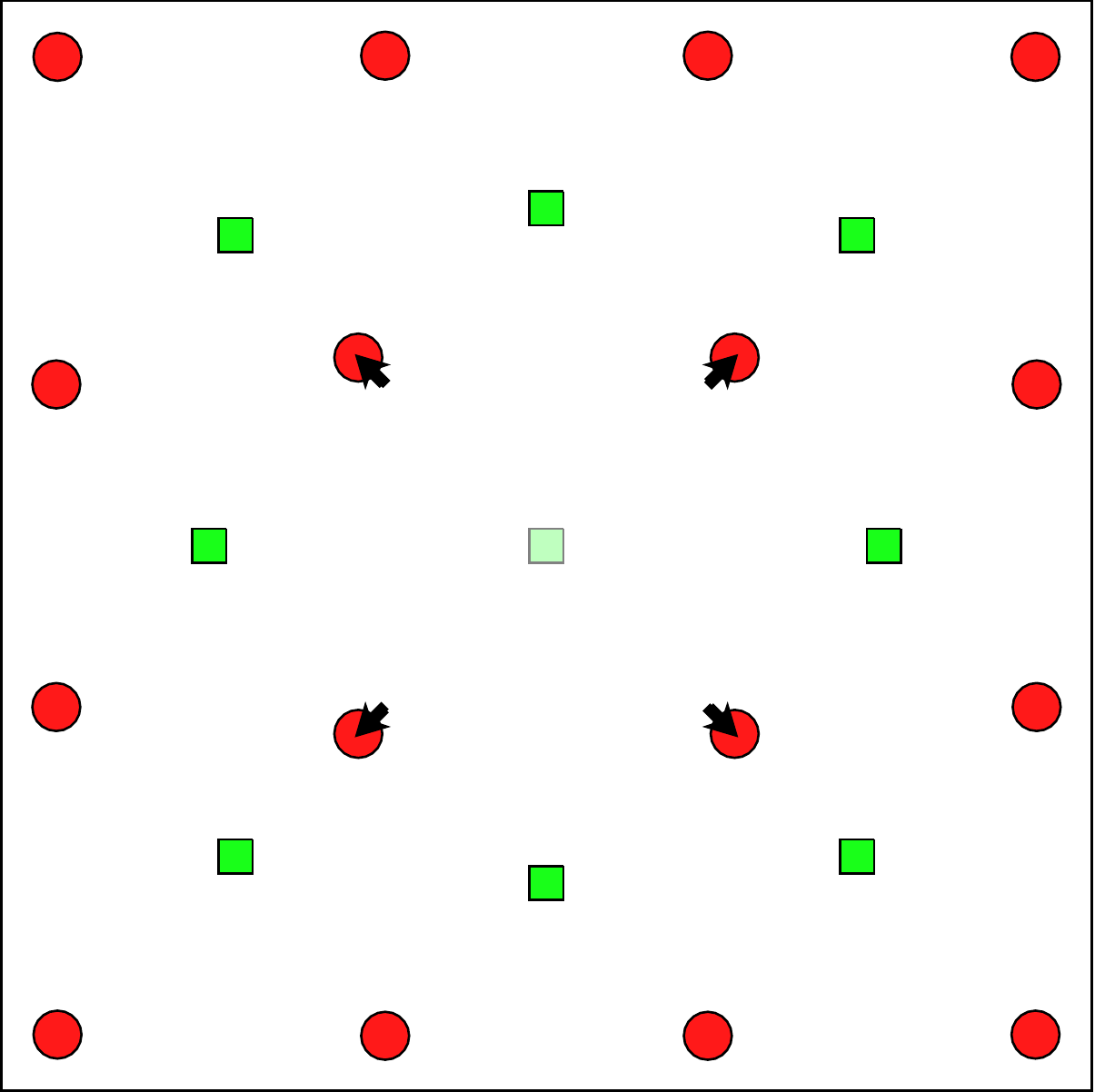}
    \caption[Local strain pattern for strontium vacancy]{Local strain
      pattern for strontium vacancy in the $\hat{e}_1 \hat{e}_2$ plane
      of strontium titanate: strontium atoms
      (\raisebox{-0.4ex}{\rlap{\textcolor[rgb]{0.1,1.0,0.1}{\FilledSquare}}{\Square}}),
      oxygen atoms
      (\raisebox{-0.4ex}{\rlap{\textcolor[rgb]{1.0,0.1,0.1}{\FilledCircle}}{\Circle}}),
      strontium vacancy
      (\raisebox{-0.4ex}{\rlap{\textcolor[rgb]{0.75,1.0,0.75}{\FilledSquare}}{\textcolor[gray]{0.60}{\Square}}}).
      Atomic displacements exaggerated by a factor of three for
      clarity (\pmb{\chemarrow}), displayed for significant
      in\hyp{}plane displacements ($> 0.1$~\AA) only.}
    \label{fig:srLocalStrain}
  \end{centering}
\end{figure}
Performing the reconstructional average gives
\begin{equation*}
  \overline{\mathbf{G}}_{\text{Sr}} = \left(
  \setlength{\arraycolsep}{2\arraycolsep}
  \begin{array}{D{.}{.}{1.2}D{.}{.}{1.2}D{.}{.}{1.2}}
    2.08 & 0.00 & 0.00\\
    0.00 & 2.08 & 0.00\\
    0.00 & 0.00 & 2.08\\
  \end{array}
  \right) \text{eV},
\end{equation*}
with a corresponding defect\hyp{}strain tensor,
\begin{equation*}
  \overline{\boldsymbol{\Lambda}}_{\text{Sr}} = \left(
  \setlength{\arraycolsep}{2\arraycolsep}
  \begin{array}{D{.}{.}{1.2}D{.}{.}{1.2}D{.}{.}{1.2}}
    1.78 & 0.00 & 0.00\\
    0.00 & 1.78 & 0.00\\
    0.00 & 0.00 & 1.78\\
  \end{array}
  \right) \text{\AA}^3.
\end{equation*}
This result expresses the tendency of the crystal to expand due to the
strontium vacancy, by an amount significantly greater than the net
effect of the oxygen vacancy.  Since the original defect defines no
unique direction, no orientational averaging is necessary, and we find
a ratio of chemical strain $\epsilon_c$ to the deviation from
stoichiometry $\delta$ in Sr$_{1-\delta}$TiO$_3$ of $\epsilon_c /
\delta = +0.030$, indicating a tendency for the crystal to expand due
to the presence of strontium vacancies.

We now examine the local strains around such a strontium vacancy after
reconstructional averaging (Figure~\ref{fig:srLocalStrain}).  Twelve
oxygen atoms (three in each of four neighboring octahedra) are nearest
neighbors to the strontium vacancy; these twelve atoms all move
directly \emph{away} from the vacancy by a distance of $0.15$~\AA.
The next\hyp{}nearest neighbors are the eight titanium atoms in each
of the surrounding octahedra, each of which moves $0.09$~\AA\ directly
\emph{toward} the strontium vacancy.  All other atoms move less than
$0.08$~\AA.

In this case, as expected, both the defect\hyp{}strain tensor and the
local strain displacements are symmetric after reconstructional
averaging.  The nearest\hyp{}neighbor displacement shows an expansion
in all directions, similar to the defect\hyp{}strain tensor.  The
next\hyp{}nearest neighbors, which have an opposite charge from that
of the nearest neighbors, move in the opposite direction, reinforcing
that there is no simple connection between local displacements and
far\hyp{}field strain patterns.

\subsection{Titanium vacancy}
\begin{figure}
  \begin{centering}
    \includegraphics[width=\columnwidth]{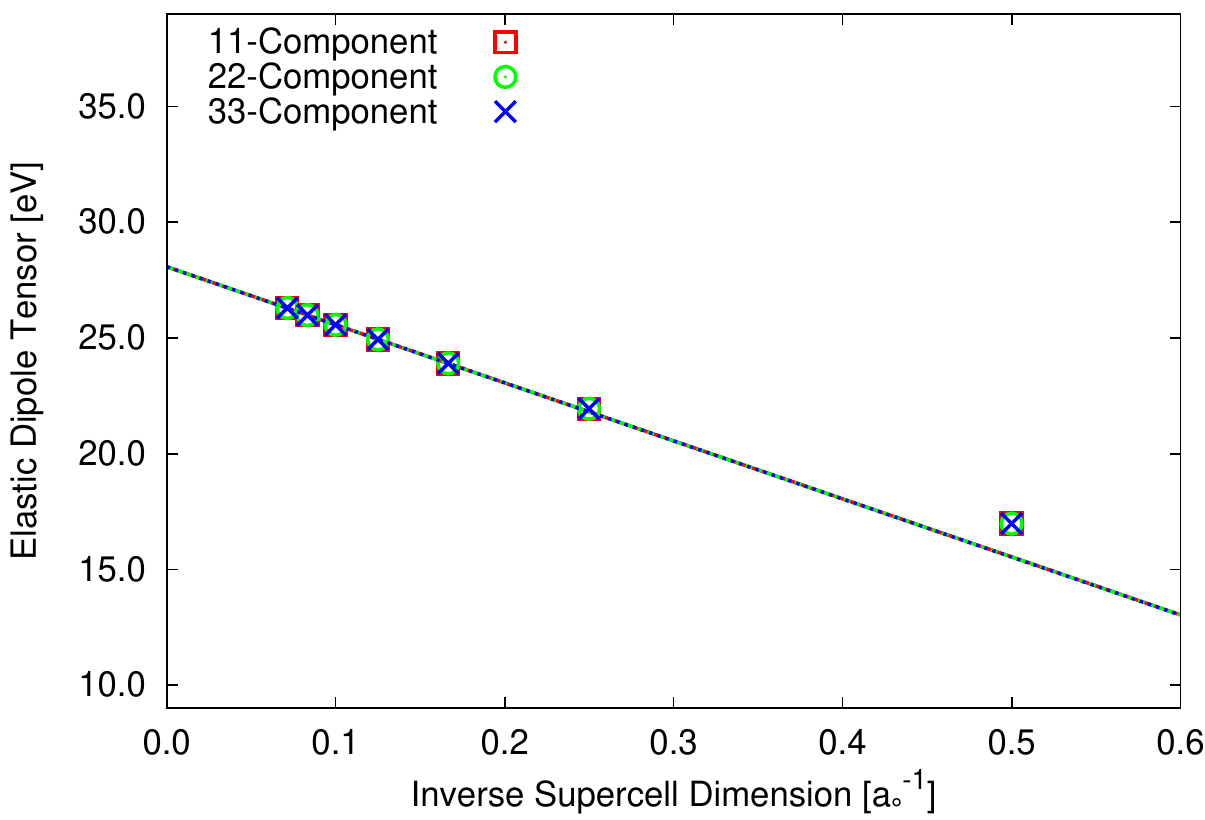}
    \caption[Elastic dipole tensor for titanium vacancy]{Diagonal
      components of elastic dipole tensor for titanium vacancy in
      strontium titanate.  Data show linear convergence with inverse
      linear dimension of supercell size.}
    \label{fig:tiElasticDipole}
  \end{centering}
\end{figure}
The titanium\hyp{}vacancy site also defines no unique direction in the
ideal crystal, and the reconstructional averaging restores the full
cubic symmetry.  For any realization of the reconstruction there are,
in fact, two distinct types of titanium site.  Each such site sits at
the center of an octahedron with either positive or negative signs of
the rotations relative to the $+[111]$ axis.  The results reported
below, prior to reconstructional averaging, correspond to the site in
which the rotation is positive.

We first report our results for the elastic dipole tensor,
\begin{equation*}
  \mathbf{G}_{\text{Ti}} = \left(
  \setlength{\arraycolsep}{2\arraycolsep}
  \begin{array}{D{.}{.}{2.2}D{.}{.}{2.2}D{.}{.}{2.2}}
    28.08 & -0.70 & -0.70\\
    -0.70 & 28.08 & -0.70\\
    -0.70 & -0.70 & 28.08\\
  \end{array}
  \right) \text{eV},
\end{equation*}
where Figure~\ref{fig:tiElasticDipole} shows the convergence of the
diagonal elements of the above tensor.
\begin{figure}
  \begin{centering}
    \includegraphics[width=\columnwidth]{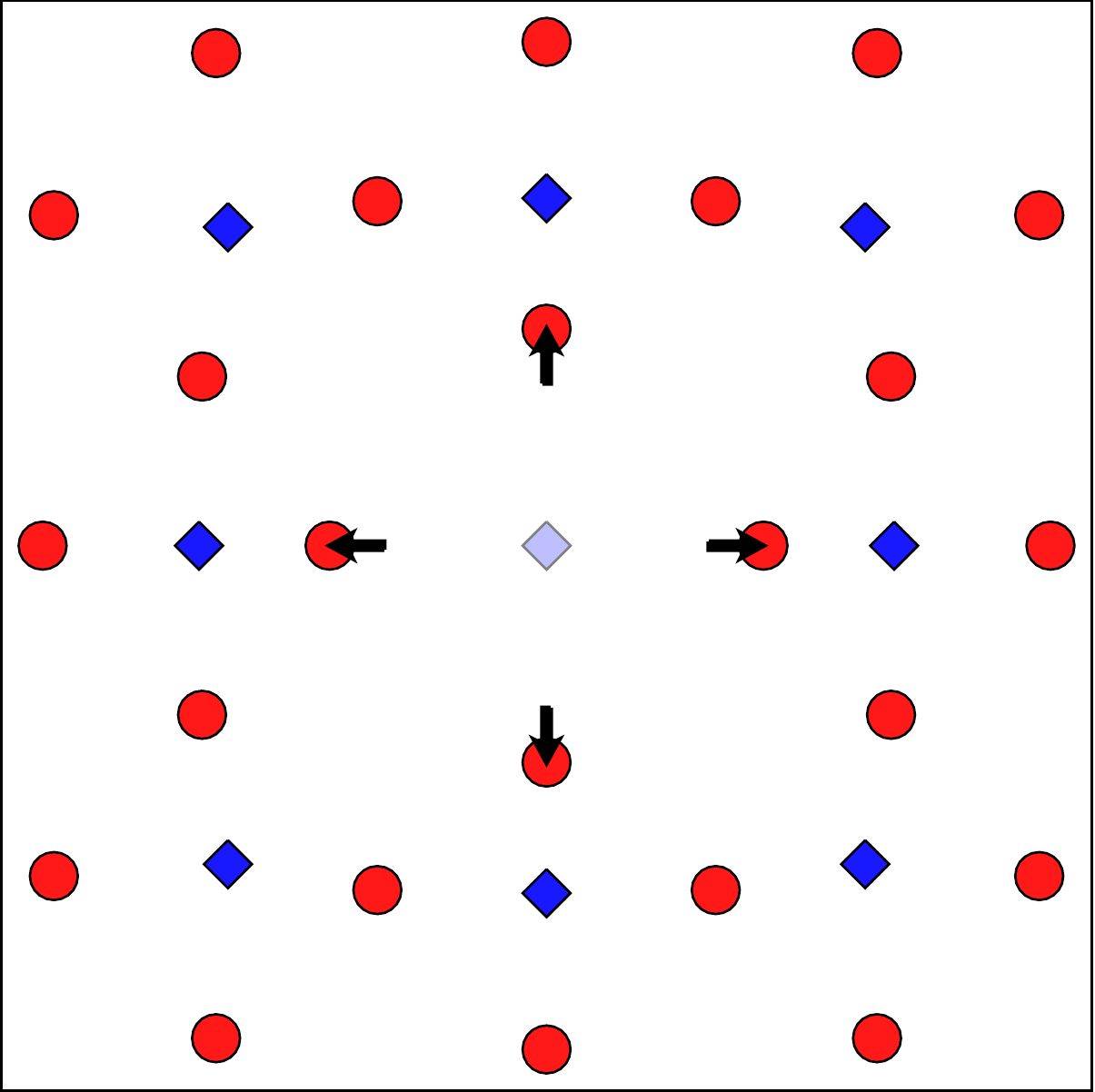}
    \caption[Local strain pattern for titanium vacancy]{ Local strain
      pattern for titanium vacancy in the $\hat{e}_1 \hat{e}_2$ plane
      of strontium titanate: titanium atoms
      (\raisebox{-0.4ex}{\rlap{\textcolor[rgb]{0.1,0.1,1.0}{\FilledDiamondshape}}{\Diamondshape}}),
      oxygen atoms
      (\raisebox{-0.4ex}{\rlap{\textcolor[rgb]{1.0,0.1,0.1}{\FilledCircle}}{\Circle}}),
      titanium vacancy
      (\raisebox{-0.4ex}{\rlap{\textcolor[rgb]{0.75,0.75,1.0}{\FilledDiamondshape}}{\textcolor[gray]{0.60}{\Diamondshape}}}).
      Atomic displacements exaggerated by a factor of three for
      clarity (\pmb{\chemarrow}), displayed for significant
      in\hyp{}plane displacements ($> 0.1$~\AA) only.}
    \label{fig:tiLocalStrain}
  \end{centering}
\end{figure}
The reconstructional average is then
\begin{equation*}
  \overline{\mathbf{G}}_{\text{Ti}} = \left(
  \setlength{\arraycolsep}{2\arraycolsep}
  \begin{array}{D{.}{.}{2.2}D{.}{.}{2.2}D{.}{.}{2.2}}
    28.08 &  0.00 &  0.00\\
     0.00 & 28.08 &  0.00\\
     0.00 &  0.00 & 28.08\\
  \end{array}
  \right) \text{eV},
\end{equation*}
with defect\hyp{}strain tensor,
\begin{equation*}
  \overline{\boldsymbol{\Lambda}}_{\text{Ti}} = \left(
  \setlength{\arraycolsep}{2\arraycolsep}
  \begin{array}{D{.}{.}{2.2}D{.}{.}{2.2}D{.}{.}{2.2}}
    23.92 &  0.00 &  0.00\\
     0.00 & 23.92 &  0.00\\
     0.00 &  0.00 & 23.92\\
  \end{array}
  \right) \text{\AA}^3.
\end{equation*}
This defect\hyp{}strain tensor expresses the tendency of the crystal
to expand due to the titanium vacancy --- a significantly greater
amount even than that of the strontium.  Again, since the original
defect defines no unique direction, no orientational averaging over
defect types is necessary.  Finally, we report a ratio of chemical
strain $\epsilon_c$ to the deviation from stoichiometry $\delta$ in
SrTi$_{1-\delta}$O$_3$ of $\epsilon_c / \delta = +0.402$, indicating a
significant tendency for the crystal to expand due to the presence of
titanium vacancies.

As depicted in Figure~\ref{fig:tiLocalStrain}, we now describe the
local strain effects on the atoms surrounding the titanium vacancy.
The nearest neighbors are six surrounding oxygen atoms which move
$0.22$~\AA\ directly \emph{away} from the titanium vacancy.  The
next\hyp{}nearest neighbors are the eight surrounding strontium atoms
along the body diagonals from the titanium vacancies; these strontium
atoms move the very significant distance of $0.52$~\AA\ directly
\emph{toward} the titanium vacancy.  The third shell is made up of six
titanium atoms separated by a lattice constant from the vacancy along
all three directions (positive and negative); all of these titanium
atoms move $0.09$~\AA\ directly \emph{away} from the titanium vacancy.
Finally, the fourth shell of atoms is comprised of twenty\hyp{}four
oxygen atoms, arranged as six groups of four oxygens, each in a
diamond\hyp{}shape with its center one lattice coordinate away from
the titanium vacancy in each lattice direction.  These oxygens each
move $0.08$~\AA\ \emph{away} from the vacancy ($0.08$~\AA\ along the
vector separating the diamond\hyp{}group from the vacancy, and
$0.03$~\AA\ along either of the two other directions, so as to cause
the diamond\hyp{}group to spread outward).  All other atoms in the
crystal move less than $0.07$~\AA.

We again observe connections between the reconstructionally averaged
local displacements and the far\hyp{}field defect\hyp{}strain tensor
which have similar symmetry.  In this case of the titanium vacancy,
the nearest\hyp{}neighbor atoms are displaced away from the vacancy,
showing the same behavior as the defect\hyp{}strain tensor.
Interestingly, however, the next\hyp{}nearest neighbors, which move
toward the titanium vacancy, actually have more than twice the
displacement of the nearest neighbors.  So here we observe that the
far\hyp{}field strain does not correlate with the largest magnitude
displacement, but instead with that of the nearest\hyp{}neighbor
atoms.

\subsection{Strontium-oxygen divacancy}
As described above in Section~\ref{sec:methods}, there are twelve
distinct orientations for the strontium\hyp{}oxygen divacancy as
defined by the V$_\text{\!Sr}$\nobreakdash-V$_\text{\!O}$ direction.
For any realization of the reconstruction there are in fact two
distinct types of strontium\hyp{}oxygen sites.  Each such strontium
site sits at the center of a cube with oxygen octahedra at the corners
with alternating signs of rotations.  The results reported below,
prior to reconstructional averaging, correspond to the
$[1\bar10\mspace{1mu}]$ defect with negative sense of rotation for the
octahedron at the $[111]$ corner.

\begin{figure}
  \begin{centering}
    \includegraphics[width=\columnwidth]{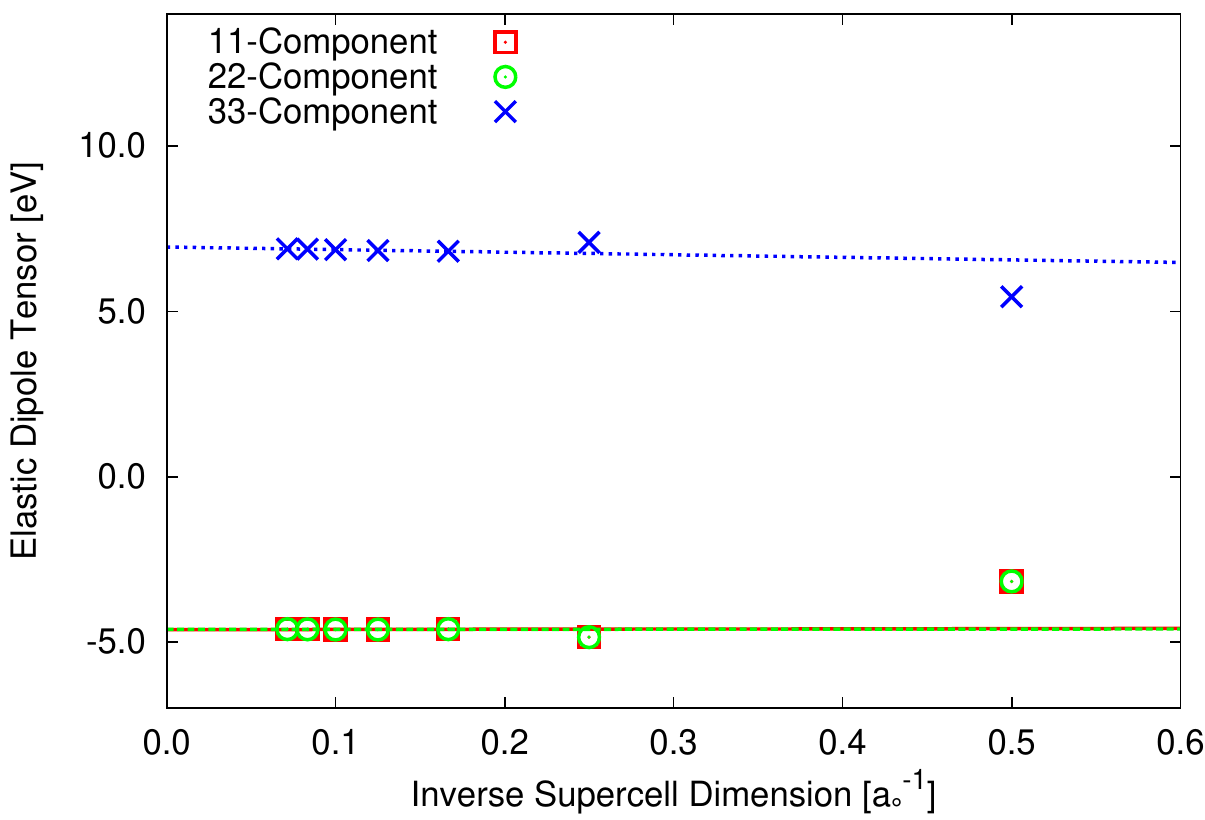}
    \caption[Elastic dipole tensor for strontium-oxygen
      divacancy]{Diagonal components of elastic dipole tensor for
      strontium\hyp{}oxygen divacancy in strontium titanate.  Data
      show linear convergence with inverse linear dimension of
      supercell size.}
    \label{fig:sroElasticDipole}
  \end{centering}
\end{figure}

The elastic dipole tensor for this defect is
\begin{equation*}
  \mathbf{G}_{\text{SrO}} = \left(
  \setlength{\arraycolsep}{2\arraycolsep}
  \begin{array}{D{.}{.}{2.2}D{.}{.}{2.2}D{.}{.}{2.2}}
    -4.62 &  3.00 & -2.28\\
     3.00 & -4.62 & -2.28\\
    -2.28 & -2.28 &  6.95\\
  \end{array}
  \right) \text{eV},
\end{equation*}
where Figure~\ref{fig:sroElasticDipole} shows the convergence of the
diagonal elements of the above tensor.  The reconstructional average
is then
\begin{equation*}
  \overline{\mathbf{G}}_{\text{SrO}} = \left(
  \setlength{\arraycolsep}{2\arraycolsep}
  \begin{array}{D{.}{.}{2.2}D{.}{.}{2.2}D{.}{.}{2.2}}
    -3.00 &  1.78 & 0.00\\
     1.78 & -3.00 & 0.00\\
     0.00 &  0.00 & 4.27\\
  \end{array}
  \right) \text{eV},
\end{equation*}
with defect\hyp{}strain tensor,
\begin{equation*}
  \overline{\boldsymbol{\Lambda}}_{\text{SrO}} = \left(
  \setlength{\arraycolsep}{2\arraycolsep}
  \begin{array}{D{.}{.}{2.2}D{.}{.}{2.2}D{.}{.}{2.2}}
    -9.37 &  2.14 &  0.00\\
     2.14 & -9.37 &  0.00\\
     0.00 &  0.00 & 17.25\\
  \end{array}
  \right) \text{\AA}^3.
\end{equation*}
This defect\hyp{}strain tensor expresses a slight tendency of the
crystal to contract due to the strontium\hyp{}oxygen divacancy.  The
lower symmetry of this defect, with its orientation along a diagonal,
leads to remaining off\hyp{}diagonal elements even after
reconstructional averaging.  However, when the above tensor is
averaged over all twelve defect orientations, the result is a constant
tensor with a uniform chemical strain per unit concentration of defect
of $-0.50$~\AA$^3$.  This corresponds to a ratio of chemical strain
$\epsilon_c$ to the deviation from stoichiometry $\delta$ in
Sr$_{1-\delta}$TiO$_{3-\delta}$ of $\epsilon_c / \delta = -0.008$,
indicating a tendency for the crystal to contract due to the presence
of strontium\hyp{}oxygen divacancies.

\begin{figure}
  \begin{centering}
    \includegraphics[width=\columnwidth]{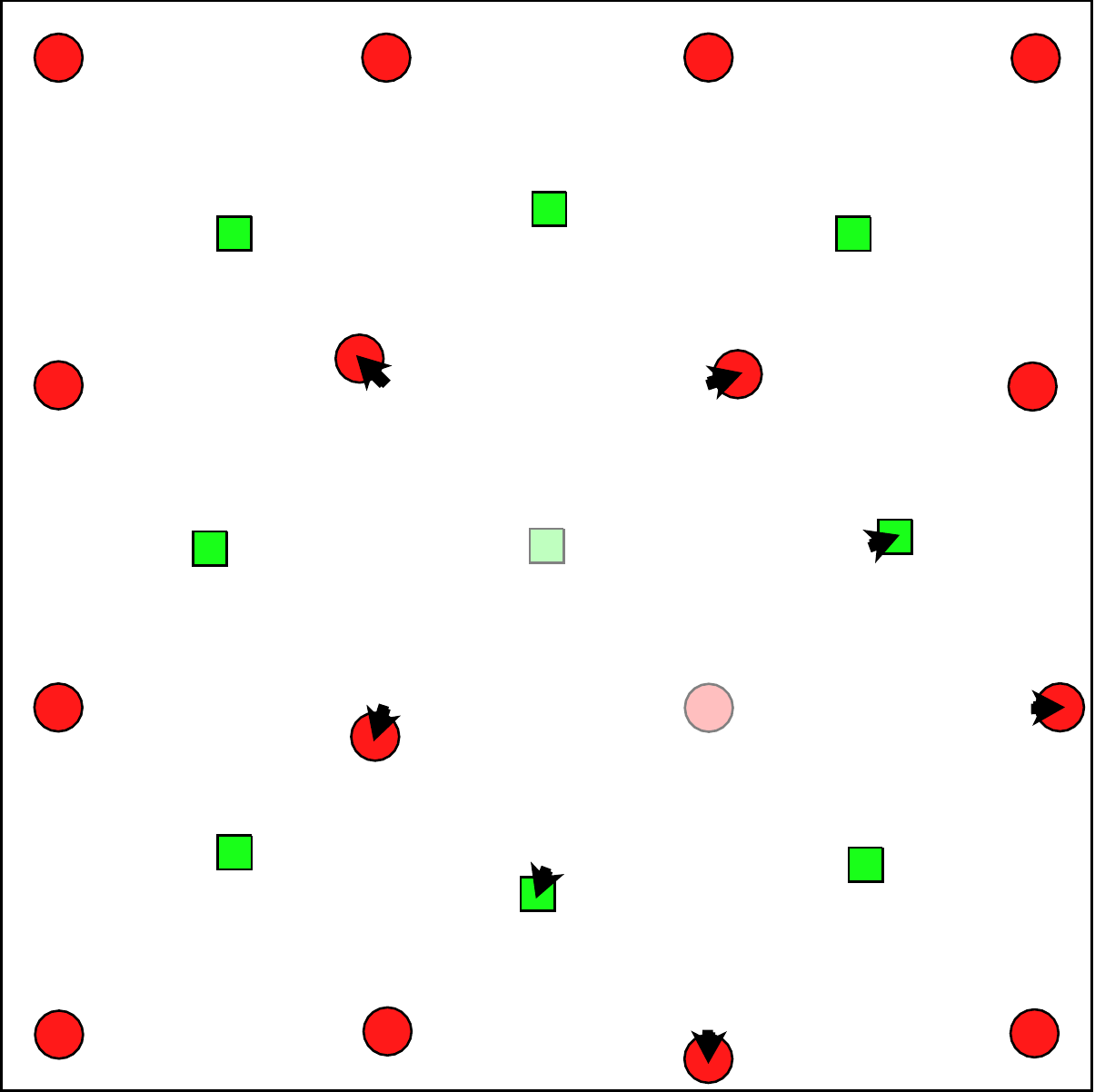}
    \caption[Local strain pattern for strontium-oxygen
      divacancy]{Local strain pattern for strontium\hyp{}oxygen
      divacancy in the $\hat{e}_1 \hat{e}_2$ plane of strontium
      titanate: strontium atoms
      (\raisebox{-0.4ex}{\rlap{\textcolor[rgb]{0.1,1.0,0.1}{\FilledSquare}}{\Square}}),
      oxygen atoms
      (\raisebox{-0.4ex}{\rlap{\textcolor[rgb]{1.0,0.1,0.1}{\FilledCircle}}{\Circle}}),
      strontium vacancy
      (\raisebox{-0.4ex}{\rlap{\textcolor[rgb]{0.75,1.0,0.75}{\FilledSquare}}{\textcolor[gray]{0.60}{\Square}}}),
      oxygen vacancy
      (\raisebox{-0.4ex}{\rlap{\textcolor[rgb]{1.0,0.75,0.75}{\FilledCircle}}{\textcolor[gray]{0.60}{\Circle}}}).
      Atomic displacements exaggerated by a factor of three for
      clarity (\pmb{\chemarrow}), displayed for significant
      in\hyp{}plane displacements ($> 0.1$~\AA) only.}
    \label{fig:sroLocalStrain}
  \end{centering}
\end{figure}

Referring to Figure~\ref{fig:sroLocalStrain}, we now examine the local
displacements after reconstructional averaging.  The situation with
this divacancy is more complicated than that of earlier isolated
atomic vacancies, leading us to characterize the atomic displacements
with respect to each independent missing atom in the
strontium\hyp{}oxygen divacancy.  The first shell is comprised of the
eleven remaining oxygen atoms that are nearest neighbors of the
strontium vacancy.  This strontium vacancy would normally have twelve
neighboring oxygen atoms forming three squares in the three planes,
each comprised of four atoms centered around the strontium vacancy;
however, one of these oxygen atoms is missing to form the divacancy.
The four oxygens, which are farthest from the oxygen vacancy but not
in the same square as the oxygen vacancy, all move \emph{away} from
the strontium vacancy by $0.14$~\AA\ ($0.10$~\AA\ along $\hat{e}_3$,
and $0.09$~\AA\ along either $\hat{e}_1$ or $\hat{e}_2$, depending
upon symmetry, with $0.004$~\AA\ along the opposite vector, chosen
such that most of each displacement is within the plane of squares to
which that oxygen belongs).  The four oxygens that are closest to the
oxygen vacancy and still not in the same square as the oxygen vacancy
also move as a group.  These oxygen atoms all move \emph{away} from
the strontium vacancy by $0.11$~\AA\ ($0.05$~\AA\ along $\hat{e}_3$,
and $0.09$~\AA\ along either $\hat{e}_1$ or $\hat{e}_2$, depending
upon symmetry, with $0.05$~\AA\ along the opposite vector, chosen such
that more of each displacement is within the plane of the square to
which that oxygen belongs).  The two oxygens, in the same square as
the oxygen vacancy and closer to such vacancy, are displaced
\emph{away} from the defect, in that plane, by $0.12$~\AA\
($0.12$~\AA\ along either $\hat{e}_1$ or $\hat{e}_2$, and $0.04$~\AA\
along the opposite vector, chosen to ensure that the displacement
maximizes its overall movement away from the remaining oxygen in this
square).  The final oxygen atom is also in the same square as the
oxygen vacancy, but is the farthest atom from the vacancy in this
square; it is displaced directly \emph{away} from the defect in that
plane by $0.14$~\AA.  Also included in the first shell are the
remaining oxygen atoms that are nearest neighbors to the missing
oxygen vacancy.  That oxygen vacancy, part of two oxygen octahedra,
has eight neighboring oxygens, four in each octahedron; of those eight
oxygens, the four closest to the strontium vacancy have already been
considered as part of the nearest neighbors to the missing strontium
atom.  The remaining four oxygen atoms in the other octahedron,
farthest from the strontium vacancy, move the appreciable distance of
$0.33$~\AA\ \emph{toward} the vector connecting the strontium and
oxygen vacancies ($0.30$~\AA\ along $\hat{e}_3$, and $0.11$~\AA\ along
either $\hat{e}_1$ or $\hat{e}_2$, depending upon symmetry, with
$0.005$~\AA\ along the opposite vector, chosen such that most of each
displacement in the $\hat{e}_1 \hat{e}_2$ plane is toward the oxygen
vacancy).  Finally, this first shell also includes the three strontium
atoms (the fourth is itself missing) nearest the oxygen vacancy.  The
one strontium atom farthest from the strontium vacancy moves directly
\emph{toward} the oxygen vacancy by $0.03$~\AA\ ($0.02$~\AA\ along
both $\hat{e}_1$ and $\hat{e}_2$).  The other two strontium atoms move
\emph{away} from the oxygen vacancy by $0.10$~\AA\ ($0.10$~\AA\ along
either $\hat{e}_1$ or $\hat{e}_2$, depending upon symmetry, with
$0.04$~\AA\ along the opposite vector, chosen such that the
displacement maximizes the distances of these strontium atoms from the
strontium vacancy).


The second shell is comprised of the eight titanium atoms that are the
nearest neighbors to the missing strontium atom.  Two of these nearest
neighboring titanium atoms, which could alternatively have been
categorized as the nearest\hyp{}neighboring shell of atoms from the
oxygen vacancy, move \emph{away} from the strontium vacancy (as well
as the oxygen vacancy) by $0.19$~\AA\ ($0.19$~\AA\ away from the
vacancies along $\hat{e}_3$, and $0.03$~\AA\ toward the strontium
vacancy along both $\hat{e}_1$ and $\hat{e}_2$).  Two other
nearest\hyp{}neighbor titanium atoms, those that are furthest from the
oxygen vacancy among these eight titanium atoms, move $0.09$~\AA\
\emph{toward} the strontium vacancy ($0.06$~\AA\ along $\hat{e}_3$,
and $0.05$~\AA\ along both $\hat{e}_1$ and $\hat{e}_2$).  The final
four of these eight nearest\hyp{}neighbor titanium atoms move
\emph{toward} the vacancies by $0.11$~\AA\ ($0.05$~\AA\ along
$\hat{e}_3$, and $0.08$~\AA\ along either $\hat{e}_1$ or $\hat{e}_2$,
depending upon symmetry, with $0.05$~\AA\ along the opposite vector,
chosen such that more of each displacement in the $\hat{e}_1
\hat{e}_2$ plane is toward the oxygen vacancy).  All other atoms move
less than $0.17$~\AA.

We now investigate correlations between these above reconstructionally
averaged local displacements and the far\hyp{}field defect\hyp{}strain
tensor.  We find surprising results in this case of the
strontium\hyp{}oxygen divacancy.  While the defect\hyp{}strain tensor
and local displacement pattern both show expansion on the
$[\mspace{1mu}001]$ axis outward from the defect, we see a
disagreement in sign in the $(\mspace{1mu}001)$ plane between the
far\hyp{}field contraction and the expansion of the nearest neighbors.
This serves as a cautionary note that far\hyp{}field and
near\hyp{}field strain patterns need not be simply related and
reinforces the importance of providing both sets of information for
further experimental analysis of x\nobreakdash-ray scattering
signatures.

\subsection{Titanium-oxygen divacancy}
As described above in Section~\ref{sec:methods}, there are six
distinct orientations for the titanium\hyp{}oxygen divacancy as
defined by the V$_\text{\!Ti}$\nobreakdash-V$_\text{\!O}$ direction.
Moreover, because of the reconstruction, there are in fact two
distinct classes of sites within each possible orientation, as
distinguished by the rotation state of the octahedron in which the
titanium sits.  Below, we report results for a $[\bar100\mspace{1mu}]$
defect with the titanium sitting in an octahedron of positive
rotation.

\begin{figure}
  \begin{centering}
    \includegraphics[width=\columnwidth]{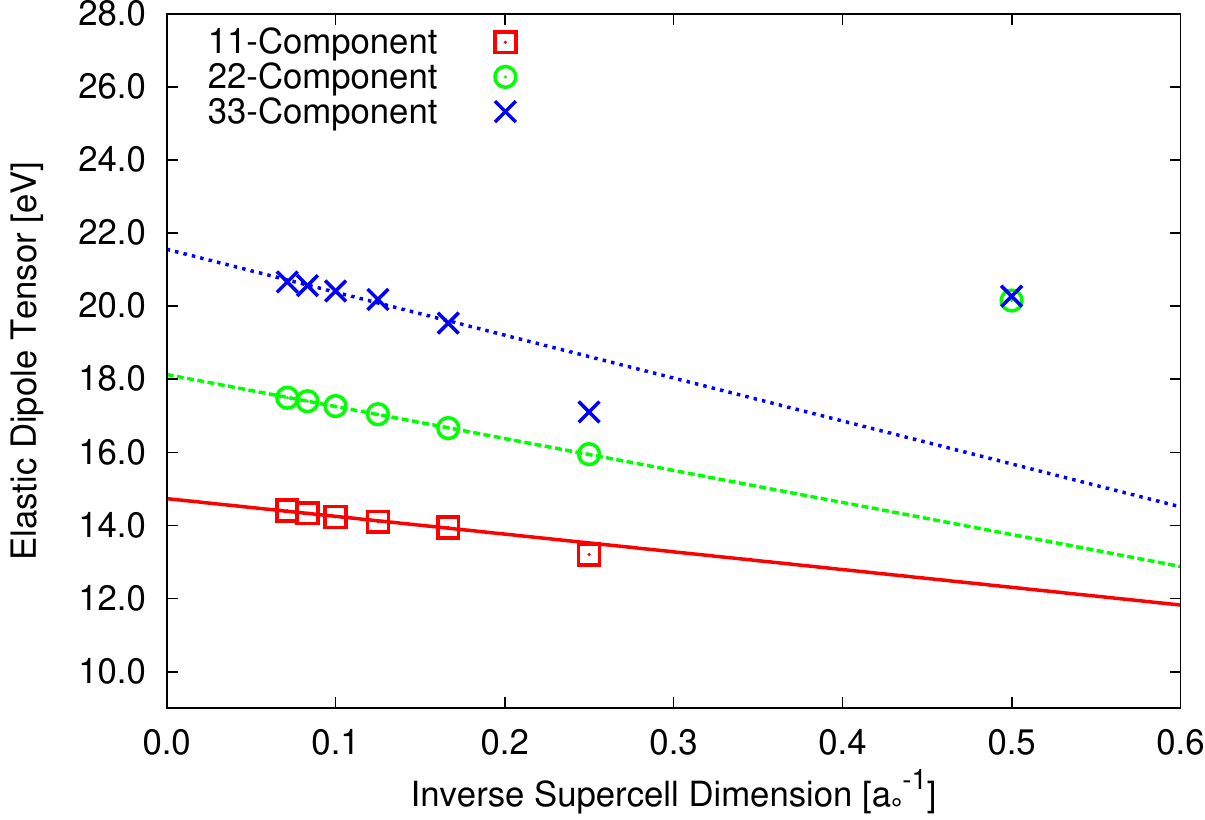}
    \caption[Elastic dipole tensor for titanium-oxygen
      divacancy]{Diagonal components of elastic dipole tensor for
      titanium\hyp{}oxygen divacancy in strontium titanate.  Data show
      linear convergence with inverse linear dimension of supercell
      size.}
    \label{fig:tioElasticDipole}
  \end{centering}
\end{figure}

The elastic dipole tensor for this defect is
\begin{equation*}
  \mathbf{G}_{\text{TiO}} = \left(
  \setlength{\arraycolsep}{2\arraycolsep}
  \begin{array}{D{.}{.}{2.2}D{.}{.}{2.2}D{.}{.}{2.2}}
    14.74 & -0.08 & -4.76\\
    -0.08 & 18.13 & -3.87\\
    -4.76 & -3.87 & 21.56\\
  \end{array}
  \right) \text{eV},
\end{equation*}
where Figure~\ref{fig:tioElasticDipole} shows the convergence of the
diagonal elements of the above tensor.
The reconstructional average gives
\begin{equation*}
  \overline{\mathbf{G}}_{\text{TiO}} = \left(
  \setlength{\arraycolsep}{2\arraycolsep}
  \begin{array}{D{.}{.}{2.2}D{.}{.}{2.2}D{.}{.}{2.2}}
    14.74 &  0.00 &  0.00\\
     0.00 & 19.84 &  0.00\\
     0.00 &  0.00 & 19.84\\
  \end{array}
  \right) \text{eV},
\end{equation*}
with corresponding defect\hyp{}strain tensor,
\begin{equation*}
  \overline{\boldsymbol{\Lambda}}_{\text{TiO}} = \left(
  \setlength{\arraycolsep}{2\arraycolsep}
  \begin{array}{D{.}{.}{1.2}D{.}{.}{2.2}D{.}{.}{2.2}}
     3.00 &  0.00 &  0.00\\
     0.00 & 21.68 &  0.00\\
     0.00 &  0.00 & 21.68\\
  \end{array}
  \right) \text{\AA}^3.
\end{equation*}
The defect\hyp{}strain tensor expresses the tendency of this defect to
expand the crystal in all directions, but primarily along the
directions orthogonal to the
V$_\text{\!Ti}$\nobreakdash-V$_\text{\!O}$ axis.  When the above
tensor is averaged over all six defect orientations (oxygen sites in
octahedra surrounding the central titanium site, recalling that the
V$_\text{\!Ti}$\nobreakdash-V$_\text{\!O}$ axis is direction
dependent), the result is a constant tensor with a uniform chemical
strain per unit concentration of defect of $+15.45$~\AA$^3$.  Such
expansion corresponds to a ratio of chemical strain $\epsilon_c$ to
the deviation from stoichiometry $\delta$ in
SrTi$_{1-\delta}$O$_{3-\delta}$ of $\epsilon_c / \delta = +0.260$,
indicating a significant tendency for the crystal to expand due to the
presence of titanium\hyp{}oxygen divacancy.

\begin{figure}
  \begin{centering}
    \includegraphics[width=\columnwidth]{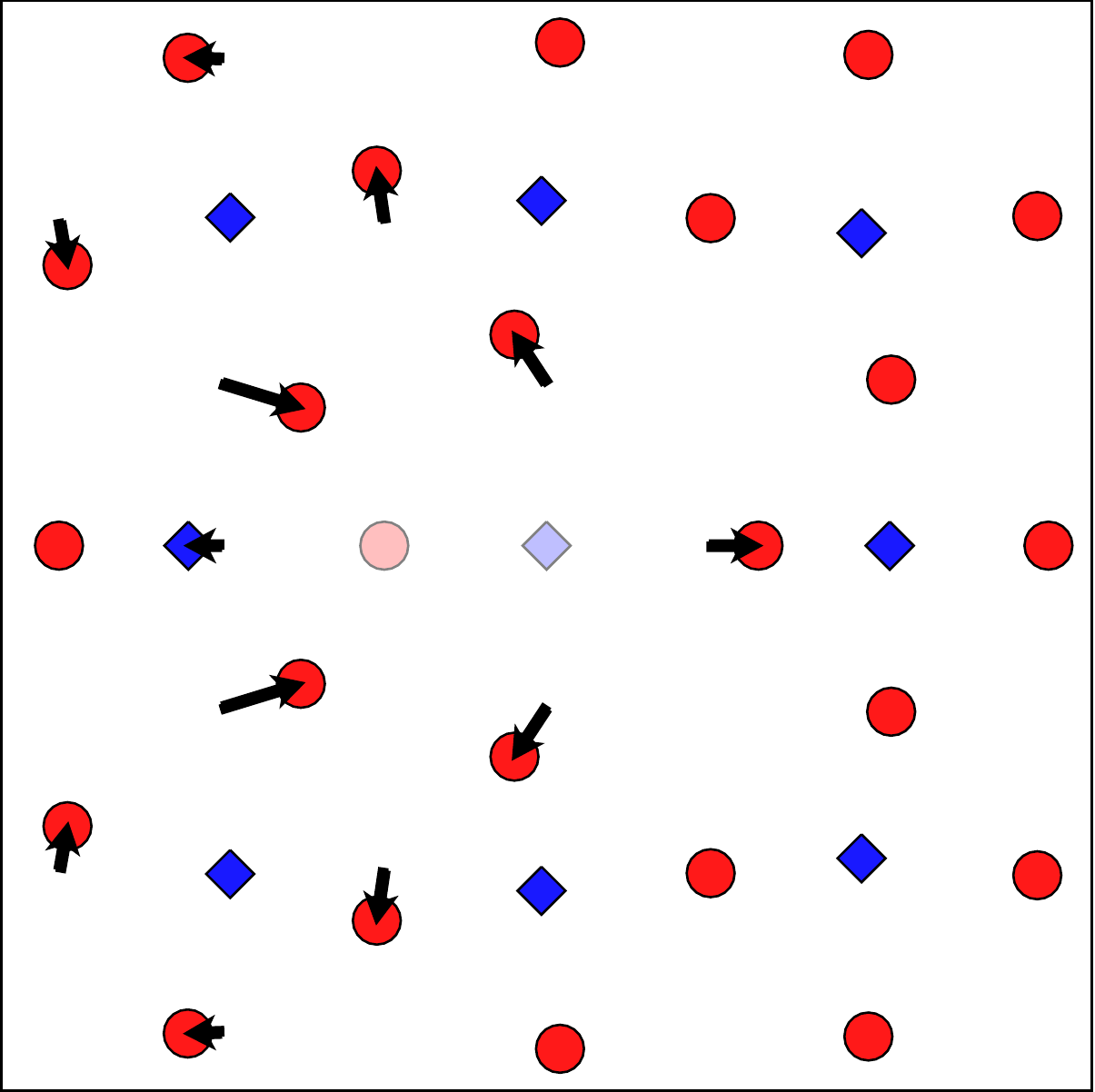}
    \caption[Local strain pattern for titanium-oxygen divacancy]{Local
      strain pattern for titanium\hyp{}oxygen divacancy in the
      $\hat{e}_1 \hat{e}_2$ plane of strontium titanate: titanium
      atoms
      (\raisebox{-0.4ex}{\rlap{\textcolor[rgb]{0.1,0.1,1.0}{\FilledDiamondshape}}{\Diamondshape}}),
      oxygen atoms
      (\raisebox{-0.4ex}{\rlap{\textcolor[rgb]{1.0,0.1,0.1}{\FilledCircle}}{\Circle}}),
      titanium vacancy
      (\raisebox{-0.4ex}{\rlap{\textcolor[rgb]{0.75,0.75,1.0}{\FilledDiamondshape}}{\textcolor[gray]{0.60}{\Diamondshape}}}),
      oxygen vacancy
      (\raisebox{-0.4ex}{\rlap{\textcolor[rgb]{1.0,0.75,0.75}{\FilledCircle}}{\textcolor[gray]{0.60}{\Circle}}}).
      Atomic displacements exaggerated by a factor of three for
      clarity (\pmb{\chemarrow}), displayed for significant
      in\hyp{}plane displacements ($> 0.1$~\AA) only.}
    \label{fig:tioLocalStrain}
  \end{centering}
\end{figure}

Now that our reconstructional averaging has symmetrized the
displacement patterns of neighboring atoms, it is instructive to
examine this set of near\hyp{}field atomic strains, shown in
Figure~\ref{fig:tioLocalStrain}.  The situation with this divacancy is
again more complicated than that of earlier isolated atomic vacancies.
There are six atoms in the first shell around the divacancy.  The five
remaining (the sixth is itself missing) oxygen atoms that are nearest
neighbors of the titanium vacancy all move \emph{away} from that
vacancy: the one that is furthest from the oxygen vacancy moves
$0.20$~\AA\ directly \emph{away} from the vacancy, while the other
four (which are in the $\hat{e}_2 \hat{e}_3$ plane) move $0.23$~\AA\
\emph{away} from the vacancy ($0.13$~\AA\ along $\hat{e}_1$ toward the
oxygen vacancy, with the remaining projection of $0.19$~\AA\ away from
the divacancy either along $\hat{e}_2$ or $\hat{e}_3$ as dictated by
the symmetry).  The sixth atom in this first shell is the sole
titanium atom that is the nearest neighbor to the oxygen vacancy; it
moves directly \emph{away} from the vacancy by $0.13$~\AA.  In the
second shell, we identify four of the oxygen atoms that are nearest
neighbor to the oxygen vacancy (we already counted the other four
nearest\hyp{}neighbor oxygens above, ``assigning'' them to the
titanium vacancy); these oxygens move $0.33$~\AA\ \emph{toward} the
oxygen vacancy ($0.31$~\AA\ along $\hat{e}_1$, with remaining
projection of $0.10$~\AA\ either along $\hat{e}_2$ or $\hat{e}_3$ as
dictated by symmetry).  Also, in this second shell, we can loosely
consider the eight nearest\hyp{}neighbor strontium atoms to the
titanium vacancy, all of which move toward the titanium vacancy: four
of these strontium atoms are in a plane that includes the oxygen
vacancy, and these move $0.22$~\AA\ \emph{toward} the titanium vacancy
($0.17$~\AA\ along $\hat{e}_1$, and $0.09$~\AA\ along both $\hat{e}_2$
and $\hat{e}_3$); the other four strontium atoms are on the opposite
side of the titanium vacancy, and these move significantly
\emph{toward} the vacancy by $0.58$~\AA\ ($0.38$~\AA\ along
$\hat{e}_1$, and $0.31$~\AA\ along both $\hat{e}_2$ and $\hat{e}_3$).
All other atoms move less than $0.20$~\AA.

We again examine connections between the reconstructionally averaged
local displacements and the far\hyp{}field defect\hyp{}strain tensor.
In this case of the titanium\hyp{}oxygen divacancy, the
nearest\hyp{}neighbor atoms to both vacancies move outward, showing
the same behavior as the defect\hyp{}strain tensor.  These atoms also
demonstrate larger movements in those directions orthogonal to the
V$_\text{\!Ti}$\nobreakdash-V$_\text{\!O}$ axis, which conforms with
the far\hyp{}field tensor.  The next\hyp{}nearest neighbors to both
vacancies move inward with significant displacements, with those
closest to the titanium vacancy moving by almost three times the
amount that the nearest\hyp{}neighbor atoms move.  So here we note
once more that the far\hyp{}field strain does not appear to correlate
with the largest magnitude displacement, but instead with that of the
nearest\hyp{}neighbor atoms.

\section{Discussion}
\label{sec:discussion}

The above results for the local strain patterns and detailed defect
tensors are now available for direct comparison with diffuse
x\nobreakdash-ray measurements; however, we are not aware of any such
x\nobreakdash-ray data to date.  Nonetheless, the fully averaged (over
both reconstructions and orientations) defect\hyp{}strain tensors
relate directly to measurements which are commonly done of chemical
strain as a function of defect concentration.
Table~\ref{tab:chemStrains} summaries the results from
Section~\ref{sec:results} for the ratios of chemical strain
$\epsilon_c$ to stoichiometric defect deviation $\delta$ for all of
the defects considered in this study.
\begin{table}
  \setlength{\doublerulesep}{0\doublerulesep}
  \setlength{\tabcolsep}{5\tabcolsep}
  \begin{tabular}{cD{.}{.}{2.3}}
    \hline\hline\\[-1.5ex]
    Defect & \multicolumn{1}{c}{$\epsilon_c / \delta$}\\[0.5ex]
    \hline\\[-1.5ex]
    V$_\text{\!O}$   &  0.001\\
    V$_\text{\!Sr}$  &  0.030\\
    V$_\text{\!Ti}$  &  0.402\\
    V$_\text{\!SrO}$ & -0.008\\
    V$_\text{\!TiO}$ &  0.260\\[0.5ex]
    \hline\hline
  \end{tabular}
  \caption[Individual ratios of chemical strain to stoichiometric
    defect deviation]{Individual ratios of chemical strain
    $\epsilon_c$ to stoichiometric defect deviation $\delta$ for
    different defects as calculated in Section~\ref{sec:results}.}
  \label{tab:chemStrains}
\end{table}

Oxygen\hyp{}vacancy concentration is widely thought to serve a crucial
role in the properties of perovskites\cite{scott2000ovo,
cherry1995oim, park1998mso, raymond1996dac, he2003fps, lo2002mro,
kimura1995fps, henrich1978sde}, is readily varied but difficult to
control\cite{tilley1977aem, szot1999sro}, and is experimentally
observed to affect chemical strain\cite{yamada1973pdi}.  Moreover,
cation stoichiometry is also difficult to
control\cite{yamamichi1994brd, taylor2003ios} and so it is uncertain
whether, as oxygen vacancies are introduced into the crystal, such
vacancies bind to cation vacancies or form in isolation.  In the
former case, where the oxygen vacancies eventually bind to preexisting
cation vacancies, the reference configuration should be the crystal
containing the cation vacancy.  Hence, it is the \emph{difference}
between the chemical strain of the oxygen\hyp{}cation divacancy and
that of the isolated cation vacancy that describes the change in the
crystal lattice as a function of varying oxygen\hyp{}vacancy
concentration.  In the latter case of isolated vacancies, the bulk
crystal is in fact the system into which these vacancies are
introduced, and thus the chemical strain as a function of
oxygen\hyp{}vacancy concentration is described precisely by that of
the isolated oxygen vacancy in our study.
Table~\ref{tab:deltaChemStrains} summarizes the resulting \emph{net}
chemical strain $\Delta \epsilon_c$ versus oxygen\hyp{}vacancy
concentration $\delta$.
\begin{table}
  \setlength{\doublerulesep}{0\doublerulesep}
  \setlength{\tabcolsep}{5\tabcolsep}
  \begin{tabular}{cD{.}{.}{2.3}}
    \hline\hline\\[-1.5ex]
    Reference state & \multicolumn{1}{c}{$\Delta \epsilon_c / \delta$}\\[0.5ex]
    \hline\\[-1.5ex]
    Bulk            &  0.001\\
    V$_\text{\!Sr}$ & -0.038\\
    V$_\text{\!Ti}$ & -0.142\\[0.5ex]
    \hline\hline
  \end{tabular}
  \caption[Net ratios of oxygen chemical strain to stoichiometric
    defect deviation]{Net ratios of chemical strains to stoichiometric
    defect deviation for oxygen vacancies, referenced against bulk and
    isolated cation vacancies as appropriate.}
  \label{tab:deltaChemStrains}
\end{table}

For the oxygen vacancy, we have the intriguing result that the elastic
dipole tensor and corresponding defect\hyp{}strain tensor almost
vanish under orientational averaging.  Thus, very little net effect on
the lattice can be expected from the presence of isolated oxygen
vacancies.  Moreover, the large anisotropy of the dipole tensor of the
oxygen vacancy and the ease of introduction and high mobility of such
vacancies should allow for the control of the population and
orientation of oxygen vacancies by applying external stress.  (For
instance, at $1$\% strain, the orientational energy differences from
$\overline{\mathbf{G}}_{\text{O}}$ are $66$~meV, or about $2.6$ times
room temperature.)  Also, such vacancies can be expected to tend to
shield internal crystalline stresses that result from materials
processing, a fact potentially related to the observed difficulties in
controlling the oxygen\hyp{}vacancy concentration during crystalline
growth.

One of the earliest sets of available experimental data on chemical
strain due to oxygen vacancies in strontium titanate comes from Yamada
and Miller\cite{yamada1973pdi}, who unfortunately found a null result.
Nonetheless, that null result places bounds which, in conjunction with
our results, allow some conclusions to be drawn.  Yamada and Miller
varied the oxygen\hyp{}vacancy concentration over a range from nearly
zero up to $3.24 \times 10^{19}$~cm$^{-3}$ ($\delta = 0.0019$ in
SrTiO$_{3-\delta}$), stating that ``no volume change upon reduction
was assumed,'' due to the experimental uncertainty of the lattice
constant ($\Delta a = 5 \times 10^{-4}$~\AA) in their
x\nobreakdash-ray diffraction measurements.  Their detection limit for
chemical strain per stoichiometric defect deviation is therefore
$\left| \epsilon_c / \delta \right| = \left| (\Delta a / a) / \delta
\right| < 0.066$, where $a$ is the cubic lattice constant.  From
Table~\ref{tab:deltaChemStrains}, it is evident that this bound is
consistent with either isolated oxygen vacancies or
strontium\hyp{}oxygen divacancies, but is inconsistent with
titanium\hyp{}oxygen divacancies.

The literature also presents studies of chemical strain due to
\emph{cation} non\hyp{}stoichiometry.  Ohnishi et
al.\cite{ohnishi2008dat} present experimental results on the ratio of
chemical strain to deviation from cation stoichiometry for samples
grown by pulsed laser deposition.  Specifically, they measure the
lattice changes for both strontium\hyp{}rich and strontium\hyp{}poor
strontium titanate.  They associate the strontium\hyp{}rich phase with
creation of Ruddlesden--Popper\cite{ruddlesden1957nco,
ruddlesden1958tcs} planar faults (extra SrO layers) and the
strontium\hyp{}poor regime with the presence of strontium vacancies,
possibly bound into strontium\hyp{}oxygen divacancies.  A
least\hyp{}squares fit to the results of Ohnishi and coworkers
provides a value for $\epsilon_c / \delta$ between $+0.13$
(non\hyp{}weighted) and $+0.14$ (weighted by reported experimental
uncertainty) for the strontium\hyp{}rich phase
(Sr$_{1+\delta}$TiO$_3$) and between $+0.5$ (non\hyp{}weighted) and
$+0.8$ (weighted by reported experimental uncertainty) for the
strontium\hyp{}poor phase (Sr$_{1-\delta}$TiO$_3$).

Our calculations of titanium\hyp{}vacancy chemical strains are not
directly relevant to the strontium\hyp{}rich phase because they do not
account for Ruddlesden--Popper planar faults.  On the other hand, our
results for strontium vacancies, under the interpretation of Ohnishi
et al.\cite{ohnishi2008dat}, should be directly relevant to their
strontium\hyp{}poor samples.  Table~\ref{tab:chemStrains} gives
$\epsilon_c / \delta = +0.030$ and $\epsilon_c / \delta = -0.008$,
respectively, for isolated strontium vacancies and bound
strontium\hyp{}oxygen divacancies.  However, both of these are an
order of magnitude smaller than the observed chemical strain.  (We
note parenthetically that, while our calculations reflect chemical
strains for isolated defects and Ohnishi and coworkers measure strains
for relatively high defect densities,
Figures~\ref{fig:srElasticDipole} and~\ref{fig:sroElasticDipole} show
that trends in our data with increasing defect concentration only tend
to reinforce our conclusions.)  We remark that our results are
consistent with the observation that, within a given structural class,
the lattice constants of the titanates are largely insensitive to the
nature of the A\nobreakdash-site cations.  (See data on
A$^{2+}$B$^{4+}$O$_3$ perovskites compiled by
Galasso\cite{galasso1969spp}.)  From both the results of our
calculations and this general observation, it seems implausible that
simple A\nobreakdash-site vacancies should produce the measured
magnitude of chemical strain.  Chemical strains of the magnitude
measured by Ohnishi and coworkers more plausibly arise from
B\nobreakdash-site vacancies or defect complexes associated with such
vacancies.

We note that, intriguingly, there is an approximate coincidence
between the order of magnitude of our calculated ratio of chemical
strain to stoichiometric deviation due to titanium vacancies, $\left|
\epsilon_c / \delta \right| = 0.4$, and the observed values ($+0.5$
and $+0.8$, depending upon weighting of fit) for the
strontium\hyp{}poor samples.  Given the magnitude of the observed
lattice expansion and the fact that \emph{only} titanium vacancies
appear capable of producing an effect of this size, we are led to the
intriguing conjecture that, perhaps, the strontium\hyp{}poor samples
exhibit defects that include titanium vacancies and thus a structure
more complex than initially thought (simple strontium vacancies or
strontium\hyp{}oxygen divacancies).  Clearly, more investigation is
needed on this point, specifically as to the nature of the point
defects in the strontium\hyp{}poor samples grown by pulsed laser
deposition.

The above measurements by Ohnishi and coworkers\cite{ohnishi2008dat}
were performed on strontium titanate samples deposited by the highly
energetic process of pulsed laser deposition\cite{willmott2000plv}.
We have learned of recent results by Brooks et
al.\cite{brooks2009ghs}, that repeat the above measurements on
strontium titanate samples grown by molecular\hyp{}beam epitaxy, which
is a lower\hyp{}energy deposition process\cite{joyce1985mbe,
stringfellow1982e} and thus less prone to the creation of point
defects\cite{willmott2000plv}.  Performing a weighted
least\hyp{}squares fit to the new data of Brooks and coworkers, we
find a value for $\epsilon_c / \delta$ of $+0.032 \pm 0.019$ in the
dilute limit of the strontium\hyp{}poor regime.  This chemical strain
ratio, published \emph{after} our initial calculated prediction was
submitted to this journal, shows good agreement with our value of
$+0.030$, and thereby lends convincing support for our methodology as
well as the applicability of empirical shell potentials to calculate
reasonable estimates of the chemical strain per stoichiometric defect
deviation of vacancies.

\section{Summary and conclusion}
\label{sec:conclusion}

We have calculated both near\hyp{} and far\hyp{}field strains for five
defects in reconstructed strontium titanate: isolated oxygen,
strontium, and titanium vacancies, as well as strontium\hyp{}oxygen
and titanium\hyp{}oxygen divacancies.  Given the propensity of the
crystal for local fluctuations in the reconstruction, we report
results both for a particular reconstructed state and as averaged over
all possible local reconstructions.  The reconstructionally averaged
near\hyp{}field strain results are presented and interpreted in terms
of the movement of neighboring shells of atoms at increasing distances
from the vacancy or divacancy.  We report far\hyp{}field strain
results in terms of both elastic dipole tensors, with and without
reconstructional averaging, and associated defect\hyp{}strain tensors,
with reconstructional averaging.  Anticipating that far\hyp{}field
effects will necessarily involve contributions from an ensemble of
defects, we also present results averaged over all possible
orientations of the defect within the bulk crystal.  From these
averaged tensors, we extract the resultant ratio of chemical strain to
stoichiometric defect deviation.  Finally, the combination of local
and long\hyp{}range results presented herein will enable determination
of x\nobreakdash-ray scattering signatures for comparison with
experimental results and should further motivate future work on defect
mechanics, including the influence of externally imposed strain (such
as in heterostructures) on vacancy populations.

For the oxygen vacancy, we find a highly anisotropic elastic dipole
tensor, with almost perfect cancellation under orientational
averaging.  This may be correlated with observed difficulties in
controlling oxygen concentration and lack of observation of effects of
oxygen\hyp{}vacancy concentration on lattice strain.  The anisotropy
of this tensor also suggests that oxygen vacancies may provide a
mechanism to shield local internal strains and that application of
external stress may allow for control of oxygen stoichiometry.  From
comparison to lattice\hyp{}strain studies, we identify both isolated
oxygen vacancies and bound strontium\hyp{}oxygen divacancies as
consistent with the experimentally observed chemical strain as a
function of oxygen\hyp{}vacancy concentration in strontium titanate.

For cation non\hyp{}stoichiometry, we find strong indications that the
point defects in strontium\hyp{}poor strontium titanate samples grown
by pulsed laser deposition are not simple strontium vacancies or
strontium\hyp{}oxygen divacancies, but likely more complicated defect
complexes.  Further, we identify indications that titanium vacancies
may play a role in these defect complexes.  Finally, during the review
process, we learned of recent experimental data, from strontium
titanate films deposited via molecular\hyp{}beam epitaxy, that show
good agreement with our calculated value of the chemical strain
associated with strontium vacancies.

\begin{acknowledgments}
  Joel Brock, Lena Fitting Kourkoutis, David Muller, and Darrell
  Schlom joined us in illuminating discussions on the experimental
  connections of these calculations.  This work was supported by the
  Cornell Center for Materials Research (CCMR) with funding from the
  Materials Research Science and Engineering Center (MRSEC) program of
  the National Science Foundation (Cooperative Agreement DMR 0520404).
\end{acknowledgments}

\bibliography{paper}

\end{document}